\documentclass[11pt]{article}
\usepackage{amsfonts}
\usepackage{amssymb}
\usepackage{graphicx}
\usepackage{setspace}
\usepackage[round]{natbib}
\usepackage{amsmath}
\usepackage{amsthm}
\usepackage{palatino}
\usepackage{paralist}
\usepackage{subfigure}
\usepackage{xcolor}
\usepackage{multirow}
\usepackage{booktabs}
\usepackage{dsfont}
\usepackage{indentfirst}
\usepackage{enumerate}
\usepackage{longtable}
\usepackage{caption}
\usepackage{rotating}
\usepackage{pdflscape}

\usepackage[hyperindex,breaklinks]{hyperref}
\hypersetup{colorlinks=true,					
  linkcolor=red,
  citecolor=blue, 							
  filecolor=magenta,							
  urlcolor=cyan								
}

\newcommand{\Var}{\text{Var}}
\newcommand{\Cov}{\text{Cov}}
\newcommand{\I}{\mathbf{I}}

\newcommand{\E}{\mathbf{E}}

\usepackage{setspace}
\title{Factor Investing: A Bayesian Hierarchical Approach\thanks{The paper was previously circulated under the title ``Factor Investing: Hierarchical Ensemble Learning". We thank Nick Polson for invaluable discussions. We also appreciate insightful comments from Victor DeMiguel, Jin-Chuan Duan, P. Richard Hahn, Fuwei Jiang (Discussant), Junye Li, Dacheng Xiu, and Jun Yu. We are also grateful for helpful comments from seminar and conference participants at ESSEC Business School, European Seminar of Bayesian Econometrics, R/Finance 2019, 2019 International Conference on FinTech in SAIF, 2019 Asia Meeting of Econometric Society, CICF 2019, and INFORMS 2019. Feng acknowledges the Strategic Research Grant from City University of Hong Kong.}}

\author{ Guanhao Feng\thanks{%
    Address: 83 Tat Chee Avenue, Kowloon Tong, Hong Kong. E-mail address:
    \texttt{gavin.feng@cityu.edu.hk}.} \\
  \textit{\small College of Business}\\
  \textit{\small City University of Hong Kong}\\
  \and  Jingyu He\thanks{%
    Address: 83 Tat Chee Avenue, Kowloon Tong, Hong Kong. E-mail address:
    \texttt{jingyuhe@cityu.edu.hk}.} \\
  \textit{\small College of Business}\\
  \textit{\small City University of Hong Kong}\\
}
\date{\today}
\begin{document}

\maketitle

\begin{abstract}
    This paper investigates the asset allocation problem when returns are predictable.
    We introduce a market-timing Bayesian hierarchical (BH) approach that adopts heterogeneous time-varying coefficients driven by lagged fundamental characteristics.
    Our approach estimates the conditional expected returns and residual covariance matrix jointly, thus enabling us to consider the estimation risk in the portfolio analysis.
    The hierarchical prior allows the modeling of different assets separately while sharing information across assets.
    We demonstrate the performance of the U.S. equity market, our BH approach outperforms most alternative methods in terms of point prediction and interval coverage. In addition, the BH efficient portfolio achieves monthly returns of 0.92\% and a significant Jensen's alpha of 0.32\% in sector investment over the past 20 years.
    We also find technology, energy, and manufacturing are the most important sectors in the past decade, and size, investment, and short-term reversal factors are heavily weighted in our portfolio. Furthermore, the stochastic discount factor constructed by our BH approach can explain many risk anomalies.
    
    
\end{abstract}

\section{Introduction}
Bayesian methods are widely applied in financial studies, ranging from stock return prediction to volatility modeling to asset allocation.\footnote{Both \cite{avramov2010bayesian} and \cite{jacquier2011bayesian} provide excellent academic surveys on Bayesian methods in financial econometrics and portfolio analysis.}
They allow researchers to incorporate prior economic beliefs in a statistical model and evaluate uncertainty in parameter estimation or even the choice of models.
Seminal Bayesian finance work, such as \cite{kandel1996predictability} and \cite{barberis2000investing}, point out that the estimation risk is non-negligible in a simple stock-bond allocation problem when stock returns are predictable. Successful deployment of the mean-variance efficient portfolio framework requires estimating conditional expected returns and their covariance matrix while accounting for estimation risk.
By contrast, traditional frequentist statistical approaches or modern machine learning predictive methods fail to properly evaluate such prediction risk.
The estimation risk issue dramatically increases when investors allocate funds among multiple assets due to the high dimension of the parameter space.

This paper introduces a Bayesian hierarchical (BH) approach to investigate the asset allocation problem when returns are predictable.\footnote{If returns are unpredictable, the mean-variance efficient portfolio is time-invariant. Therefore, the econometric interest lies in the estimation property of unconditional expected returns and the covariance matrix.}
In particular, we model returns by market-timing macro predictors and assume lagged fundamental characteristics drive their predictor strength (regression coefficients).
Our approach estimates regression coefficients and the covariance matrix jointly, thus making possible the creation of the mean-variance efficient portfolio.
The benefits of our approach are twofold.
First, Bayesian methods allow one to model expected returns and the covariance matrix jointly and learn the uncertainty of the estimation.
Second, our BH approach fits multiple assets separately while sharing data information through the hyperparameters of a hierarchical prior distribution. This data sharing feature improves individual asset fitness while maintaining model heterogeneity for different assets.

The empirical exercise aims to allocate funds across multiple risky assets based on market-timing macro predictors and fundamental characteristics. The investment universe includes sector portfolios, tradable risk factors, and characteristics-sorted portfolios. We are interested in sector or factor rotation under different macroeconomic conditions. Notably, we project time-varying coefficients of each asset onto its fundamental characteristics. This setup is similar to \cite{avramov2006predicting}, but they instead assume coefficients of all asset returns are driven by the same macroeconomics predictors. Consequently, the conditional expected returns and covariance matrix are driven by both macro predictors and fundamental characteristics; thus, we can study factor rotation under changing macroeconomic conditions over time. Furthermore, the mean-variance efficient portfolio created by the BH approach serves as a stochastic discount factor, and it can explain most of the known asset pricing anomalies.

The rest of the paper is organized as follows.
We first introduce the econometric and financial motivation in section \ref{sec: motivation_e} and \ref{sec: motivation_f}, and report the methodology and empirical overviews in section \ref{sec: overview_m} and \ref{sec: overview_e}.
Section \ref{sec: literature} reviews relevant literature in Bayesian econometrics and empirical asset pricing.
Section \ref{sec: method} introduces our BH approach setting, the corresponding Markov chain Monte Carlo (MCMC) sampling algorithm, and the portfolio construction scheme.
Section \ref{sec: empirical} documents the empirical performance for asset return prediction, portfolio evaluations, and asset pricing implications.
Section \ref{sec: conclusion} concludes the paper.

\subsection{Motivation: Hierarchical Modeling} \label{sec: motivation_e}

Documenting asset return predictability is a popular research topic among both academic researchers and practitioners. The existing empirical literature focuses on testing the existence of return predictability for a single market index (e.g., S\&P 500) using market-timing macro predictors, such as dividend yield, treasury bill rates, inflation, and so on. However, some researchers argue most of these predictors are unstable or even spurious \citep{welch2008comprehensive}. The forecasting power of these predictors undoubtedly varies substantially for different assets, and the current literature struggles to find consistent predictability evidence across \textit{multiple} assets \citep{huang2020time}.

Particular attention should be paid to modeling multiple assets. In this case, the time series of asset return is usually short (e.g., monthly data with hundreds of observations), but the number of time series is relatively large (e.g., hundreds or even thousands of assets). Researchers either give up the potential benefit from massive data and model each asset independently (time series modeling) or ignore the heterogeneity of assets, stack all time series together, and train a single model (pool modeling).\footnote{For example, \cite{gu2020empirical} use pooled modeling, whereas \cite{huang2020time} use time series modeling.}
Neither way properly takes advantage of information from massive data. Time series modeling has poor individual asset fitness due to the small sample size being used, whereas pooled modeling is naive and loses heterogeneous signals. Stock return data usually suffer from extremely low signal-to-noise ratios, and the predictive power of predictors changes under different macroeconomic conditions. Therefore, improving model fitness is critical.

Our BH approach sheds light on this problem from a Bayesian perspective. BH modeling provides a useful framework for studying the joint predictability for multiple assets. The hierarchical model\footnote{Chapter 5 of \cite{rossi2012bayesian} is an excellent textbook reference for hierarchical modeling.} provides a way to share information across numerous assets. Hence, it can handle the cross-sectional return dependence. The sampling process of our BH approach can be interpreted as two intuitive cyclic steps.
The first step (information sharing step) takes condition on the hierarchical prior and updates each asset's regression coefficients separately. Note the hierarchical prior provides information from all other assets.
The second step (information grouping step) evaluates regression coefficients of all asset returns and updates parameters of the hierarchical prior to share information. Intuitively, the hierarchical prior builds an overline bridge linking multiple asset returns. The within-asset part describes the predictor-return dynamics, and the cross-asset part incorporates the heterogeneity of predictor existence and strength.

Despite the advantages above, our Bayesian framework is flexible enough to model heterogeneous time-varying coefficients driven by each asset's lagged fundamental characteristics.
\cite{avramov2006predicting} find these macro predictors are linked to the underlying business cycles for conditional investing or market timing, and their analysis focuses on individual stocks.
Our paper pursues a similar goal: to assess the economic value of predictability and show portfolio strategies successfully rotate across different risk factor styles during changing business conditions. Furthermore, our BH approach models the covariance matrix simultaneously, thus enabling the creation of mean-variance efficient portfolios.

\subsection{Motivation: Factor Investing} \label{sec: motivation_f}

Thousands of stocks are traded on the secondary market in the U.S., beyond what investors would like to invest in. Academic researchers and practitioners generally perform dimension reduction on the vast individual stock universe.
They usually study the asset allocation problem in a small number of tradable portfolios instead of thousands of individual stocks.
\cite{kozak2020shrinking} provide a Bayesian shrinkage method to estimate the efficient portfolio weights for the stochastic discount factor (SDF) in empirical asset pricing.

Sorting is arguably the most popular approach to create portfolios among financial economists.
Researchers sort stocks based on values of one fundamental characteristic, such as market equity values, book-to-market ratios, and so on. Then, they split the continuum of stocks into several consecutive groups and create portfolios using stocks within each group (i.e., characteristics-sorted portfolios).\footnote{For example, some index funds or ETFs classify and group individual stocks into mega-cap, mid-cap, and small-cap portfolios, which is equivalent to sorting individual stocks by market equity values.}
This sorting mechanism\footnote{\cite{feng2020deepchar} discuss the security-sorting mechanism used in asset pricing.} only uses a simple relative ranking of firms in the characteristics and does not consider the scale nor build statistical models.
Many papers document that stocks within the same group behave similarly, and the characteristics-sorted portfolios have relatively stable dynamic relations with macro predictors and fundamental characteristics.

Moreover, many characteristics-sorted portfolios are shown with relatively monotonic average returns, which implies return predictability through past characteristics.
The long-short portfolio (long top and short bottom sorted portfolios) returns are treated as a proxy for an underlying risk factor or an anomaly \citep{fama1993common}.
The current trend of factor investing is driven by these intuitive innovations of dimension reduction in the vast stock universe. Investors can directly invest in a small number of risk factors or the characteristics-sorted portfolios.
However, the enlarging zoo of risk factors \citep{feng2020taming} creates a new challenge of selecting risk factors.

Our BH approach joins this literature branch by considering investing in many risk factors or characteristics-sorted portfolios to further reduce the dimension of the investing universe. We create mean-variance efficient portfolios based on our estimation of conditional expected returns and the covariance matrix.
The cross-sectional asset pricing literature generally tests characteristics-sorted portfolios or risk factors.
We follow this path to introduce the hierarchical structure for return predictability across multiple assets.
Moreover, the mean-variance efficient portfolio constructed by our method can be viewed as the SDF when considering a broad cross section of assets. We demonstrate in section \ref{sec: empirical} that it can explain many known anomalies.

\subsection{Overview of Methodology} \label{sec: overview_m}

Our paper bridges the gap between Bayesian econometrics and empirical finance. The highlights are as follows.
First, we propose a predictive system for the cross section of asset returns using market-timing macro predictors, considering parameter uncertainty.
Second, each asset's predictive model adopts heterogeneous time-varying coefficients driven by each asset's lagged fundamental characteristics.
Third, the hierarchical prior structure allows the sharing of information while maintaining heterogeneity for different assets.
Finally, our model estimates conditional expected returns and covariance matrix of assets jointly, thus making it possible to create mean-variance efficient portfolios.

The primary goal of our approach is to create a common predictor-return dynamic and study the optimal portfolio choice for $N$ risky assets\footnote{We mainly consider the tangency portfolio optimization in the portfolio analysis.}. Let $r_{i,t+1}$ denote the excess returns and let $x_t$ represent the vector of lagged macro predictors. Our predictive regressions provide a system for cross-sectional return prediction.
\begin{equation*}
    r_{i,t+1} = \alpha_{i,t} + x_{t}^\intercal\beta_{i,t} + \epsilon_{i,t+1},  \quad \forall i = 1,\cdots, N.
\end{equation*}
We assume $\alpha_{i,t}$ and $\beta_{i,t}$ are driven by characteristics and have a hierarchical prior distribution.
We follow the recent development of machine learning in asset pricing, such as \cite{gu2020empirical} and \cite{feng2020deep}, and model asset returns by combining macro predictors and fundamental characteristics. The advantage of our BH approach is that each asset uses information not limited to its time series but borrows information from others. By contrast, all other prediction studies face a trade-off between sample size and model heterogeneity. A full Bayesian evaluation considers the parameter uncertainty for interval prediction and efficient portfolio optimization where machine learning fails.

\subsection{Overview of Empirical Study} \label{sec: overview_e}

We perform various empirical exercises in the U.S. equity market using monthly data from 1978 to 2018.
With 10 market-timing macro predictors and 20 representative fundamental characteristics, we test three representative groups of portfolios: 10 sector portfolios (Sector 10), 20 long-short factors (Factor 20), and 100 characteristics-sorted portfolios (Char-Sort 100). A complete list of the factors and portfolios is presented in the appendix \ref{app: char}.
These three groups of portfolios have a diverse representation of the cross section.
The purpose of the increasing number of assets is to evaluate our approach's robustness to a higher dimension asset universe.

In terms of asset return prediction, we find our BH approach outperforms many strong benchmark models, including Lasso, random forest (RF), principal component regression (PCR), and the Bayesian method in \cite{avramov2006predicting} (AC2006).
Although the Bayesian forecast is biased, our BH approach performs slightly weaker than the moving average, but is still robust.
Furthermore, our BH approach has more accurate prediction interval coverage than alternatives because it estimates the covariance matrix jointly and considers the estimation risk of coefficients.

For efficient portfolio performance, we also find substantial positive gains using our BH approach.
Our BH approach outperforms many robust benchmarks in terms of risk-adjusted evaluation, including the naive equally weighted portfolio, the passive investing portfolio (S\& P 500), and the predictive methods listed above.
In the past 20 years, our BH approach in sector investment delivers average monthly returns of 0.92\% and a significant Jensen`s alpha of 0.32\%.
We find technology, energy, and manufacturing are the most important sectors in the past decade for sector rotation. For factor rotation, size, investment, and short-term reversal factors have been heavily weighted in the efficient portfolio.
Finally, the SDF constructed by our BH approach explains mail anomalies that alternative models fail to.

\subsection{Related Literature} \label{sec: literature}
The body of literature on Bayesian methods in asset pricing is extensive and long established.\footnote{Early studies include \cite{shanken1987bayesian}, \cite{harvey1990bayesian}, and \cite{mcculloch1991bayesian}. Though not directly related, another body of Bayesian learning papers in asset pricing can be a future research direction for our models, such as \cite{lewellen2002learning}, \cite{johannes2014sequential}, and \cite{fulop2015self}} 
\cite{pastor2000portfolio} and \cite{pastor2000comparing} introduce Bayesian asset pricing models to the portfolio optimization problem.
\cite{avramov2004stock} add time-varying alphas and time-varying risk premia into Bayesian asset pricing models, and \cite{avramov2006asset} show explanations for the size and value effects through time-varying coefficients.
We learn a similar structure from them and use lag characteristics and macro predictors to drive the time-varying coefficients while incorporating a novel, seemingly unrelated regression (SUR) model with hierarchical priors. Though the lens of factor investing, our paper identifies useful factors through their efficient portfolio weights.

Bayesian methods have also appeared sporadically in the literature of portfolio analysis.
For the simple stock-bond allocation problem, \cite{kandel1996predictability} and \cite{barberis2000investing} focus on posterior predictive distribution of returns for the expected returns and covariance matrix, and \cite{avramov2006predicting} extend the model by allowing time-varying coefficients, which are driven by characteristics.
Unlike analyzing predictive distributions, our approach draws samples from the posterior by the MCMC approach.
\cite{polson2000bayesian} introduce a Bayesian seemingly unrelated regression (SUR) model with hierarchical prior and employ informative priors on the expected returns and covariance matrix. But they model the covariance matrix separately rather than learning it through the SUR model. We follow a similar SUR structure. In addition, we allow dynamic regression coefficients and learning the covariance matrix jointly with coefficients.

Finally, our paper is related to the recent literature on high-dimensional cross-sectional asset pricing models.\footnote{Other recent related publications include \cite{lettau2020PCA}, \cite{demiguel2020portfolio}, and \cite{freyberger2020dissecting}.}
\cite{gu2020empirical} investigates machine learning algorithms to forecast asset returns by many firm characteristics and macro predictors.
\cite{feng2020deep} also find strong predictability evidence through a benchmark combination using characteristics-sorted portfolio returns.
\cite{kelly2019characteristics} adopt a characteristics-driven time-varying coefficient model for principal component analysis.
Our study is similar to this literature in predicting returns by high-dimensional firm characteristics and macro predictors, but we propose a novel statistical model. In addition, we go one step further, from return prediction to portfolio optimization.

\section{Methodology} \label{sec: method}
Section \ref{sec: predictive} discusses the model settings. Section \ref{sec: sur} illustrates the reformulation of the likelihood as an SUR model. Section \ref{sec: prior} discusses the hierarchical prior specification. The MCMC scheme is presented in section \ref{sec: mcmc}. Finally, section \ref{sec: portfolio} shows portfolio optimization based on information extracted by the MCMC algorithm.

\subsection{Bayesian Predictive Model} \label{sec: predictive}
The goal of a Bayesian investor is to maximize portfolio performance though a Bayesian predictive model of asset returns. Suppose he or she observes the historical return $R_t = (r_{1,t}, \cdots, r_{N,t})$ of $N$ assets, $P$ asset characteristics $z_t$ for each asset, and $Q$ market-timing macro predictors $x_t$. An investor updates his or her portfolio regularly, at time period $t$, models the joint predictive distribution of returns $f(R_{t+1} \mid D_t)$, and calculates asset allocation weight $W_t = (w_t^1, \cdots, w_t^N)$ accordingly. In the next time period, the realized return of the portfolio is
\begin{equation}
    R^{p}_{t+1} = W_{t}^\intercal R_{t+1}.
\end{equation}
We denote all historical data observed up to period $t$, including asset returns $R_t$, characteristics $x_t$, and macro predictors $z_t$ as $D_t = (R_t, x_t, z_t)$. The asset allocation decision is made at the end of period $t$ based on the information in $D_t$.

If returns are unpredictable, the efficient portfolio can be constructed using the unconditional mean and covariance matrix of returns. If returns are predictable, one needs to learn the source and mechanism of return predictability.
We follow the Bayesian predictive regression model in \cite{kandel1996predictability} but adopt a conditional predictive formula with time-varying coefficients. For each asset $i$, its return at time period $t+1$ is assumed to be
\begin{equation}\label{eqn: predictive}
    r_{i,t+1} = \alpha_{i,t} + x_{t}^\intercal\beta_{i,t} + \epsilon_{i,t+1},
\end{equation}
where $x_{t}$ is the vector for $Q$ macro predictors. The residual vector $(\epsilon_{1,t+1}, \cdots, \epsilon_{N,t+1})^\intercal$ contains shocks to all asset returns and is assumed to follow a multivariate normal distribution $N(0, \Sigma)$ with a dense covariance matrix $\Sigma$.

Conditional modeling with time-varying coefficients is one solution for those ``seemingly useless" predictors debated in \cite{welch2008comprehensive}. The regression coefficients $\alpha_{i,t}$ and $\beta_{i,t}$ are assumed to be time-varying, driven by asset characteristics as follows:
\begin{equation}\label{eqn: alpha}
    \begin{aligned}
        \alpha_{i,t} & = \eta^{a}_{i} + z_{i,t}^\intercal \theta^{a}_{i}, \\
        \beta_{i,t}  & = \eta^{b}_{i} + \theta^{b}_{i}z_{i,t},
    \end{aligned}
\end{equation}
where $\theta^{b}_{i}$ is a matrix coefficient of size $Q\times P$, and $z_{i,t}$ is the vector for $P$ portfolio characteristics. \cite{avramov2004stock} gives a similar time-varying coefficients setup but assumes a common factor structure for all assets, and factor loadings and alphas are driven by lagged stock characteristics. If we plug the time-varying coefficients (\ref{eqn: alpha}) into equation (\ref{eqn: predictive}), we obtain an unconditional predictive regression in terms of $z_t$, $x_t$, and their interactions $z_t \otimes x_t$:
\begin{equation}\label{eqn: uncond}
    r_{i,t+1} =  \eta^{a}_{i} + z_{i,t}^\intercal \theta^{a}_{i} +  x_{t}^\intercal\eta^{b}_{i} + \left(x_{t}\otimes z_{i,t}\right)^\intercal \theta^{b}_{i} + \epsilon_{i,t+1},
\end{equation}
where $\otimes$ stands for the Kronecker product.
Moreover, our linear Bayesian model allows for more intuitive results over black-box machine learning predictive models.

Notice equation (\ref{eqn: uncond}) models the $i$-th asset by macro predictors $x_t$ and its own characteristics $z_{i,t}$. Instead of modeling each asset separately, we prefer group modeling to estimate covariance and share information across different assets, which we achieve by SUR \citep{zellner1962efficient} with the hierarchical prior on coefficients. The SUR models allow joint modeling of asset returns and the covariance matrix. The hierarchical prior provides a channel to share information across different assets while modeling each asset's return by its characteristics. \cite{polson2000bayesian} provide a similar hierarchical SUR setting, but they model the covariance matrix by a separate shrinkage estimator.
In the following two subsections, we present our SUR setup and the hierarchical prior.

\subsection{Seemingly Unrelated Regressions}\label{sec: sur}
Before discussing the hierarchical model, we simplify the notations of equation (\ref{eqn: uncond}) as
\begin{eqnarray}
    r_{i, t+1} &=& f_{i, t}^\intercal b_i + \epsilon_{i,t+1},
\end{eqnarray}
where $f_{i, t} = [1, z_{i,t}, x_t, (x_t\otimes z_{i,t})]$ indicates all regressors, symbol $\otimes$ denotes the Kronecker product, and $b_i = [\eta_i^\alpha, \theta_i^\alpha, \eta_i^b, \theta_i^b]$ is a vector of all corresponding regression coefficients. For each asset $i$, we stack the equations of different time periods $t$ as
\begin{eqnarray}
    r_i &=& f_i^\intercal b_i + \epsilon_i,
\end{eqnarray}
where $r_i = (r_{i,2}, \cdots, r_{i,T+1})^\intercal$, $\epsilon_i = (\epsilon_{i,2}, \cdots, \epsilon_{i,T+1})^\intercal$, and $f_i$ is a matrix with $T$ rows.

The system of all $i = 1, \cdots, N$ assets is written as an SUR setup as follows. Stacking all equations, we have
\begin{eqnarray}
    R &=& FB + E,
\end{eqnarray}
where
\begin{equation*}
    R = \begin{bmatrix}r_1 \\ r_2 \\ \vdots \\ r_N\end{bmatrix}, \quad F = \begin{bmatrix}
        f_1 & 0      & 0      & 0   \\
        0   & f_2    & 0      & 0   \\
        0   & \cdots & \cdots & 0   \\
        0   & 0      & 0      & f_N
    \end{bmatrix},\quad
    B = \begin{bmatrix}b_1 \\ b_2 \\ \vdots \\ b_N\end{bmatrix}, \quad
    E = \begin{bmatrix} \epsilon_1 \\ \epsilon_2 \\ \vdots \\ \epsilon_N   \end{bmatrix}.
\end{equation*}
Here, $R$ is an $NT \times 1$ vector of a stacked vector of firm returns, $F$ is an $NT \times NK$ block diagonal matrix, $B$ is an $NK \times 1$ vector, and $E$ is an $NT\times 1$ stacked vector of residuals.

Estimating the covariance matrix of many assets, $\Sigma$, is a challenging problem in general. The literature of portfolio optimization mostly develops a covariance matrix estimator independently from return predictions. Researchers either assume a low-dimensional factor structure or shrink the empirical estimator. Our SUR setup allows for joint modeling for asset returns and the covariance matrix.

We assume the covariance matrix of the stacked residual $E$ has the form $\text{Cov}(E) := \Omega = \Sigma \otimes I_N$, where $\Sigma$ is a dense matrix of cross-sectional covariance, and $I_N$ is an $N\times N$ identity matrix. This structure implies assets are correlated in the cross section but are independent across different periods, which is a standard assumption in the empirical asset pricing literature. This specification differs from the standard SUR model in \cite{zellner1962efficient} as well as \cite{polson2000bayesian}, who do not assume any constraint of $\Omega$ and estimate the $NT\times NT$ matrix directly. Therefore, we reduce the dimension of the parameter space dramatically from $N^2T^2$ to $N^2$.

Estimating the dense covariance matrix $\Sigma$ enables us to create the mean-variance portfolio. Many popular machine learning algorithms, such as Lasso, random forest, or deep learning, cannot model this common shock jointly with returns but assume a constant variance over time or across assets.

\subsection{Hierarchical Prior Specification}\label{sec: prior}
The current empirical literature mainly focus on the time-series predictability evidence for index returns, such as the S\&P 500. Similar to the field of cross-sectional asset pricing, researchers are also interested in the common predictors across assets, such as individual stocks, sector portfolios, and sorted portfolios on market equities.

To study the joint predictability across assets, rather than putting an independent prior on $b_i$ for each asset,  we assume a hierarchical structure with a common prior mean. Suppose $b_i$ has an independent and identical normal prior
\begin{eqnarray}
    b_i &\sim& N(\bar{b}, \Delta_b) \quad\forall i = 1, \cdots, N,
\end{eqnarray}
where $\bar{b}$ is the prior mean for $b_i$, and $\Delta_b$ is a dense prior covariance matrix. Furthermore, the prior mean and covariance are assumed to have a second layer prior:
\begin{eqnarray}
    \bar{b} &\sim& N(\bar{\bar{b}}, \Delta_{\bar{b}}), \\
    \Delta_b &\sim& IW(\nu_b, V_b).
\end{eqnarray}
This setup is the standard normal-inverse-Wishart conjugate prior for $\bar{b}$ and $\Delta_b$.  It assumes the average predictability of all predictors is $\bar{\bar{b}}$, and $\Delta_b$ represents the confidence of this predictability belief. We set  $\bar{\bar{b}} = 0$ in practice, implying the belief a priori that the average predictability of those predictors is zero. Finally, this prior specification contains four groups of hyperparameters: $\{\bar{\bar{b}}, \Delta_{\bar{b}}, \nu_b, V_b\}$.

The normal prior on $b_i$ is equivalent to $L_2$ shrinkage of coefficients similar to the ridge regression. Because $\Delta_{\bar{b}}$ is a parameter to learn from the data, our model can adapt the shrinkage level based on the information from the data. This ability is the primary advantage of our BH approach over alternatives. When many potential and weak predictors exist, the hierarchical prior can properly shrink coefficients. From the frequentists' perspective, simply testing $H_0: \bar b_j = 0$ reveals the joint predictability implication of the jth predictor.

Notice the dimension reduction assumption $\Omega = \Sigma \otimes I_N$, where we assume the standard inverse-Wishart prior on $\Sigma$ with two hyperparameters $\nu_\Sigma$, and $V_\Sigma$ as
\begin{equation}
    \Sigma \sim IW(\nu_\Sigma, V_\Sigma).
\end{equation}
The MCMC sampler in section \ref{sec: mcmc} updates $\Sigma$, and $\Omega$ can be recovered by simple calculation.

The model likelihood function is multivariate normal:
\begin{equation}
    l(E\mid B, \Omega) \propto |\Omega|^{-\frac{1}{2}}\exp\left\{-\frac{1}{2}(R - FB)^\intercal \Omega^{-1}(R - FB)\right\}.
\end{equation}
Hence, the joint posterior distribution can be expressed as
\begin{equation}
    p(B, \Omega, \bar{b}, \Delta_b \mid R,F) \propto l(E\mid B, \Omega) p(\Omega) p(B\mid \bar{b}, \Delta_b)p(\bar{b})p(\Delta_b).
\end{equation}
Based on the above prior and likelihood specification, the next subsection describes the corresponding Gibbs sampler for model inference.

\subsection{Markov Chain Monte Carlo Scheme} \label{sec: mcmc}

The design of our Gibbs sampler is straightforward because all priors are conjugate to the likelihood. As discussed in section \ref{sec: sur}, we assume $\Omega = \Sigma \otimes I_N$ and update smaller $\Sigma$ instead of $\Omega$. The corresponding Gibbs sample is
\begin{itemize}
    \item[(1)]  Update $B$ through a multivariate normal distribution (information sharing step):
          \begin{equation}
              B\mid \bar{b},\Delta_b, \Omega, R, F \sim N\Big(b^{*}, (F^\intercal \Omega^{-1}F + I_N\otimes \Delta_b^{-1})^{-1}\Big),
          \end{equation}
          where
          \begin{equation}
              b^{*} = \left(F^{\intercal}\Omega^{-1}F  + I_N\otimes \Delta_b^{-1}\right)^{-1} \left(F^{\intercal}\Omega^{-1}R + (I_N\otimes \Delta_b^{-1})(\mathbf{1}_N \otimes \bar{b})\right).
          \end{equation}
          $I_N$ is an $N\times N$ dimensional identity matrix, and $\mathbf{1}_N$ denotes an $N \times 1$ vector of ones.

    \item[(2)] Update $\bar{b}$ through a multivariate normal distribution (information grouping step):
          \begin{equation}
              \bar{b} \mid B, \Delta_b \sim N\left(\left(\Delta_{\bar{b}}^{-1}  + N \Delta_b^{-1}\right)^{-1}\left(\Delta_{\bar{b}}^{-1}\bar{\bar{b}} + \Delta_b^{-1}(\mathbf{1}_N \otimes I_K)^{\intercal} B\right), \left(\Delta_{\bar{b}}^{-1}  + N \Delta_b^{-1}\right)^{-1}\right),
          \end{equation}
          where $\bar{\bar{b}}$ is set to 0 for the purpose of shrinkage.


    \item[(3)] Update $\Delta_b$ through an inverse-Wishart distribution:
          \begin{equation}
              \Delta_b\mid B, \bar{b} \sim IW\Big(\nu_b + N, ((D - \bar{b}\otimes \mathbf{1}_N^\intercal)(D - \bar{b}\otimes \mathbf{1}_N^\intercal)^\intercal + V_b) )^{-1}\Big).
          \end{equation}

    \item[(4)] Update $\Sigma$ through an inverse-Wishart distribution:
          \medskip
          \begin{equation}
              \Sigma \mid B, R, F\sim IW(\nu_\Sigma + T, V_\Sigma + \tilde{E}^\intercal \tilde{E}),
          \end{equation}
          where $\tilde{E} = [\hat{\epsilon}_1, \cdots, \hat{\epsilon}_N]$ is a $T\times N$ matrix of residuals, $\hat{\epsilon}_i = r_i - f_i^\intercal b_i$. Calculate $\Omega = \Sigma \otimes I_N$ accordingly.
\end{itemize}

Convergence is always a concern for MCMC algorithms. We confirm the convergence of the algorithm by examining the trace plot of posterior samples and calculating the effective sample size (ESS) for all parameters using the R package \texttt{coda} \citep{plummer2006coda}. The trace plot stabilizes quickly, and we achieve an average effective sample size of 1,800 out of 2,000 posterior draws, which indicates that the sampler explores the posterior efficiently. See the appendix \ref{app: mcmc} for details.

\subsection{Predicting Returns and Creating Efficient Portfolio}\label{sec: portfolio}
One can simply plug the Bayesian estimates $b_i$ and new data of $z_t, x_{i,t}$ into equation (\ref{eqn: uncond}) to make predictions for the next time period. We take the posterior average as our estimation for parameters. Specifically, for the kth draw in a total of K MCMC samples, we have
\begin{equation}
    \widehat r^{(k)}_{i,t+1} = f_{i,t}^\intercal b^{(k)}_i.
\end{equation}
The conditional expected returns and covariance matrix proceeds similarly:
\begin{equation}
    \begin{aligned}
        \E(R_{t+1} \mid D_t) & = \frac{1}{K}\sum^K_{k=1} \widehat r^{(k)}_{i,t+1}   \\
                             & = f_{i,t}^\intercal \frac{1}{K}\sum^K_{k=1}b^{(k)}_i
    \end{aligned}
\end{equation}
\begin{equation}
    \begin{aligned}
        \Cov(R_{t+1} \mid D_t) & =  \Cov\left( \widehat r_{i,t+1}, \widehat r_{j,t+1}\right)                                                                                                                              \\
                               & =  \Cov\left( f_{i,t}^\intercal b_{i,t+1}, f_{j,t}^\intercal b_{j,t+1}\right)                                                                                                            \\
                               & =  f_{i,t}^\intercal \left( \frac{1}{K}\sum^K_{k=1} b^{(k)}_{i} b^{(k)\intercal}_{j} - \frac{1}{K}\sum^K_{k=1} b^{(k)}_{i} \frac{1}{K}\sum^K_{k=1} b^{(k)\intercal}_{j} \right) f_{j,t}.
    \end{aligned}
\end{equation}
Notably, our BH approach is able to provide more information about the uncertainty of the parameters. For example, we can calculate the interval forecast for the vector of returns, or value at risk (VaR), based on the posterior variance of parameters. By contrast, most machine learning methods fail to estimate the covariance matrix and only give the point prediction of returns.

Hence, given the predicted return and covariance matrix, we create a mean-variance efficient portfolio by optimizing the standard utility function.
Simply plug in the Bayesian estimates of $\E(R_{t+1} \mid D_t)$ and $\Cov(R_{t+1} \mid D_t)$ for optimal portfolio weight calculation. The portfolio is built to maximize the mean-variance utility function:
\begin{equation} \label{eqn: utility}
    U(W_t) = \exp\left\{ \E(R_{p,t+1}) - \frac{\gamma}{2}\Var(R_{p,t+1}) \right\},
\end{equation}
where $R_{p,t+1} = W_t^\intercal R_{t+1}$ is the return of the portfolio with allocation weight $W_t$ and
\begin{equation}
    \E(R_{p,t+1}) = W_t \E(R_{t+1} \mid D_t), \quad \Var(R_{p,t+1}) = W_t^{\intercal} \Cov(R_{t+1} \mid D_t) W_t.
\end{equation}
The optimal weight is
\begin{equation}\label{eqn: optimal_weight}
    W_t^{*} = \arg\max_{W_t} U(W_t).
\end{equation}
The risk-aversion parameter $\gamma$ indicates an investor's risk preference. One convenient property for this optimal portfolio is
\begin{equation}
    W_t \propto \Cov(R_{t+1}\mid D_t)^{-1}\E(R_{t+1}\mid D_t).
\end{equation}
This formula is the same portfolio-efficiency estimate as \cite{kozak2020shrinking}, who adopt a maximum a posteriori (MAP) estimator. They consider incorporating a shrinkage prior to shrink the cross section of risk anomalies: factors and characteristics-sorted portfolios. The efficient portfolio generated through a large cross section is supposed to be the stochastic discount factor that prices the cross section. Our approach can tell the same story. One major difference is that ours provides a full Bayesian estimate that allows for posterior inference, whereas the MAP estimate fails.

In the empirical exercise section, we place some constraints on the optimization of equation (\ref{eqn: optimal_weight}). We restrict short selling, leverage and assume full position, or
\begin{equation}\label{eqn: optimal_weight_constraint}
    \begin{aligned}
        W_t^{*} & = \arg\max_{W_t} U(W_t)                                 \\
                & \text{s.t.} \sum_{i=1}^N W_t^{i} = 1, \quad W_t \geq 0.
    \end{aligned}
\end{equation}
This constraint eliminates leverage and short selling. We only consider the allocation of risky assets; thus, we assume the investor spends every dollar on assets. We observe macro predictors and fundamental characteristics in the current period. Given that fundamental characteristics drive time-varying predictor coefficients, our framework is a convenient way to provide or update the one-step-ahead or multi-step-ahead optimal asset allocation weights. We illustrate this convenient property in the empirical analysis.

\section{Empirical Study} \label{sec: empirical}
Section \ref{sec: data} introduces details of the data. The portfolio rebalancing scheme is listed in section \ref{sec: metrics}. Section \ref{sec: forecasting} illustrates the prediction performance comparison between our BH method and other workhorse benchmarks. In section \ref{sec: portfolio_result}, we further compare the performance of the efficient portfolio and add the predictor usefulness evaluation. Finally, we show the results of asset pricing implications in section \ref{sec: ap}.

\subsection{Data Sample} \label{sec: data}
Our data sample consists of extensive coverage of the U.S. equity market. The data are from January 1978 to December 2018. We test our approach on three major groups of assets, including 10, 20, and 100 portfolios. First, we follow the Fama-French 10-industry classifications (sector 10).\footnote{Fama and French form 10 sector portfolios at the end of June of year t with the Compustat SIC codes for the fiscal year ending in the calendar year t-1.} It classifies thousands of individual stocks into 10 portfolios. By analyzing this group of assets, we can also learn the impact of sector rotation through time. We also investigate the predictability of 20 risk factors (Factor 20) and 100 characteristics-sorted portfolios (Char-Sort 100), which are commonly used testing assets in asset pricing studies.

The number of selected portfolios ranges from 10 to 100 in the exercise. It helps us test our BH approach's robustness to different cross-section sizes for portfolio construction. Furthermore, these three asset groups are constructed in fundamentally different ways for the stock universe. It also allows us to assess the robustness of our BH approach to different types of portfolios.

We take 20 portfolio characteristics $z_{i,t}$ from the six major categories: momentum, value, investment, profitability, frictions (or size), and intangibles. Appendix \ref{app: char} lists details of all 20 characteristics. The data are from the public access library of \cite{hou2020replicating}. Some minor differences exist because our work focuses on monthly asset return prediction. First, we modify the characteristics formulas, so they are updated monthly. Second, the characteristics-sorted portfolios are $5 \times 1$ univariate-sorted\footnote{We also follow the NYSE breakpoints and work on the same stock universe as Fama-French three factors.} every month. Third, the risk factors are long-short portfolios on the corresponding characteristics-sorted portfolios (top-bottom or bottom-top).

Following \cite{kelly2019characteristics}, We standardize the cross section of the monthly characteristics of firms in the range of $[-1,1]$.\footnote{For example, the market equity in 2018 December is uniformly standardized to $[-1, 1]$. The firm with the lowest market equity is -1, and the firm with the highest market equity is 1. Every month, a ``size" factor longs firms with size $< -0.6$ and shorts firms with size $\geq 0.6$. Therefore, this uniform standardization is a non-standard standardization that transforms the data onto $[-1, 1]$ every month. If a firm has missing values for some characteristics, the imputed values are 0, which implies the firm is not important in security sorting.} To calculate portfolio characteristics, we take the average of the sector's standardized characteristics and characteristics-sorted portfolios. For long-short risk factors, we take spread differences between the long and short portfolio characteristics.

Firm characteristics and market-timing macro predictors applied in this paper are the same as in \cite{feng2020deep}. Those 10 macro predictors are listed in Appendix \ref{app: macro}. \cite{welch2008comprehensive} study return prediction of S\&P 500 returns using market-timing predictors, and \cite{amihud2002illiquidity} also constructs the market illiquidity for the S\&P 500 index for market timing. \cite{gu2020empirical} use eight predictors of \cite{welch2008comprehensive} in their analysis, all of which are covered in our list.

\subsection{Portfolio Rebalancing}\label{sec: metrics}
We estimate the model using a rolling window of the past 21 years. The prediction period is from January 1999 to December 2018.
Given the constraint of computational resources, we only update the model annually (calendar year).
Each year, the model is trained using a rolling window of the past 21 years (252 months).
The model is fixed for the incoming year, but regression coefficients and predictors are updated monthly.
Thus, we can provide an updated monthly forecast for expected returns.
The dynamic asset allocation procedure is based on re-estimating and then rebalancing portfolios based on the updated monthly forecasts.
We apply the same rolling-window strategy for other methods in the empirical comparison.

Below are the step-by-step details of our design of the dynamic portfolio optimization.
\begin{enumerate}
    \item We estimate the model using a rolling window of the past 21 years (252 months) of historical data annually. For the predictive model at year $K$, we use the monthly data from year $(K-21)$ to $(K-1)$ as the training data.\footnote{Appendix \ref{app: mcmc} shows diagnostic plots to check convergence of the MCMC algorithm.} we draw 3,000 posterior samples and throw away the first 1,000 draws as burn-in periods.

    \item When predicting asset returns in year $K$, we plug in the lagged monthly predictors $x_t$ and $z_t$. The model is fixed for the entire year $K$ because it is reestimated annually, but the regression coefficients $\alpha_{i,t}$ and $\beta_{i,t}$ in equation \ref{eqn: alpha} update according to $z_{t}$. The covariance matrix is an output from the Bayesian model estimation, which is fixed for the entire year $K$.

    \item Other methods (Lasso, RF, PCR) do not provide estimates for the covariance matrix. Therefore, we use the sample covariance matrix of fitted residuals instead.

    \item We optimize the mean-variance utility and update portfolio weights every month, with constraints of long-only and no leverage.
\end{enumerate}
We compare our BH method with other commonly used machine learning methods, including Lasso, RF, and PCR. The tuning parameters of these approaches are selected by threefold cross-validation with a comprehensive list of parameter candidates. Every fold of validation contains seven consecutive years in the rolling window setting and addresses business cycles' impact on model selection. Additionally, we compare our approach with \cite{avramov2006predicting} (AC2006)\footnote{AC2006 has a problem inverting matrices when the matrix of predictors is singular, due to their uninformative prior specification. Their model does not allow for heterogeneous predictors, $z_{i, t}$ for each asset. Therefore, we implement their model only with our 10 macro predictors. Specifically, we use the first five predictors in Appendix \ref{app: macro} as their equity predictors, and the second five as macroeconomic predictors.}.

\subsection{Forecasting Evaluation}\label{sec: forecasting}
To gauge the performance of our Bayesian hierarchical approach, we compare it with alternative methods in terms of a few key metrics, including out-of-sample R-square, coverage, and length of the predictive interval. The measurements are defined as follows:
\begin{eqnarray}
    R_{OOS}^2 &=& 1 - \frac{\sum_{i,t}(\widehat R_{i,t} - R_{i,t})^2}{\sum_{i,t}(\bar R_{i,t} - R_{i,t})^2}, \\
    C_{OOS} &=& \frac{1}{N\times T}\sum_{i,t}\I\left( \widehat R_{i,t} - 1.96\times\widehat \sigma_{i,t} < R_{i,t} < \widehat R_{i,t} + 1.96\times\widehat \sigma_{i,t}\right), \\
    IL_{OOS} &=& 2\times1.96\times \sigma_{i,t},
\end{eqnarray}
where $\I(\cdot)$ is the indicator function. We follow the literature and use moving average $\bar R_{i,t}$ as the point prediction benchmark in $R_{OOS}^2$ calculation. The out-of-sample predictive interval is at the 95\% level (from 2.5\% to 97.5\% quantiles of the posterior distribution).

Note our BH approach offers a dynamic heterogeneous predictive interval for each asset, so we take the average interval across time and assets. Additionally, our Bayesian predictive interval is wider due to the uncertainty in parameter estimation. By contrast, most machine learning methods cannot estimate the variance or covariance of assets. Thus, we instead calculate the residual variance or covariance. Table \ref{tab: prediction} presents results of the predictions. The table lists results of two different prior settings (tight vs. mild) \footnote{The two prior settings are mild, $\bar{\bar{b}} = 0, \Delta_{\bar{b}} = \text{diag}(0.1, K), \nu_b = 1001 + K, V_b = \text{diag}(3, K)$; tight, $ \bar{\bar{b}} = 0, \Delta_{\bar{b}} = \text{diag}(0.1, K), \nu_b = 5001 + K, V_b  = \text{diag}(3, K)$. Note the mean of inverse Wishart distribution $IW(\nu,V)$ is $V / (\nu - p - 1)$ where $p$ is the dimension. Here, the tight prior implies a small prior mean of $\Delta_b$; thus, the prior standard deviation of $b$ is around 0.02, corresponding to stronger regularization on $b$. } for the BH method, indicating strong shrinkage and relatively flat prior parameter settings, respectively.

The first panel of table \ref{tab: prediction}  summarizes results for the past 20 years. We find the BH approach has a moderately higher $R_{OOS}^2$ than most other methods, including Lasso, RF, PCR, and AC2006. Admittedly, the slightly negative $R_{OOS}^2$ implies the BH approach does not outperform the moving average in the overall sample. However, if we focus only on the recent decade in the bottom panel, both the point and interval predictions of the BH approach are better than most others.

Additionally, the BH approach gives the best out-of-sample predictive interval coverage in the overall sample and subsamples. Notably, the better accurate interval coverage is due to wider intervals or considering the uncertainty of parameter estimation.
The pattern is robust across all three types of portfolios: 10 sectors, 20 factors, and 100 characteristics-sorted portfolios. These are the empirical facts for the importance of accounting for estimation risk in return prediction.

\subsection{Portfolio Performance}\label{sec: portfolio_result}
Next, we proceed to dynamic asset allocation. We mainly study the risk-adjusted performance of portfolios and the asset pricing implication in this section. We compare the average monthly return, annualized Sharpe ratio, and Jensen's alpha of the monthly updated optimal portfolio using different methods. We also include the $1/N$ naive diversification strategy (equally weighted portfolio, or EW), and the buy-and-hold strategy (SPY, exchange-traded fund of S\&P 500 index) as passive investment benchmarks. All of the active investing portfolios are updated monthly using the corresponding forecasts of conditional expected returns and covariance matrix.\footnote{To limit the impact of transaction fees, the monthly portfolio turnover rate is restricted to be smaller than 50\%, and the maximum position of a single asset is 50\% of the total portfolio.}

\begin{figure}[ht]
    \begin{center}
        \caption{Cumulative Portfolio Performance}\label{fig: cumreturn}
        \vspace{-0.35cm}
        \includegraphics[width = \textwidth]{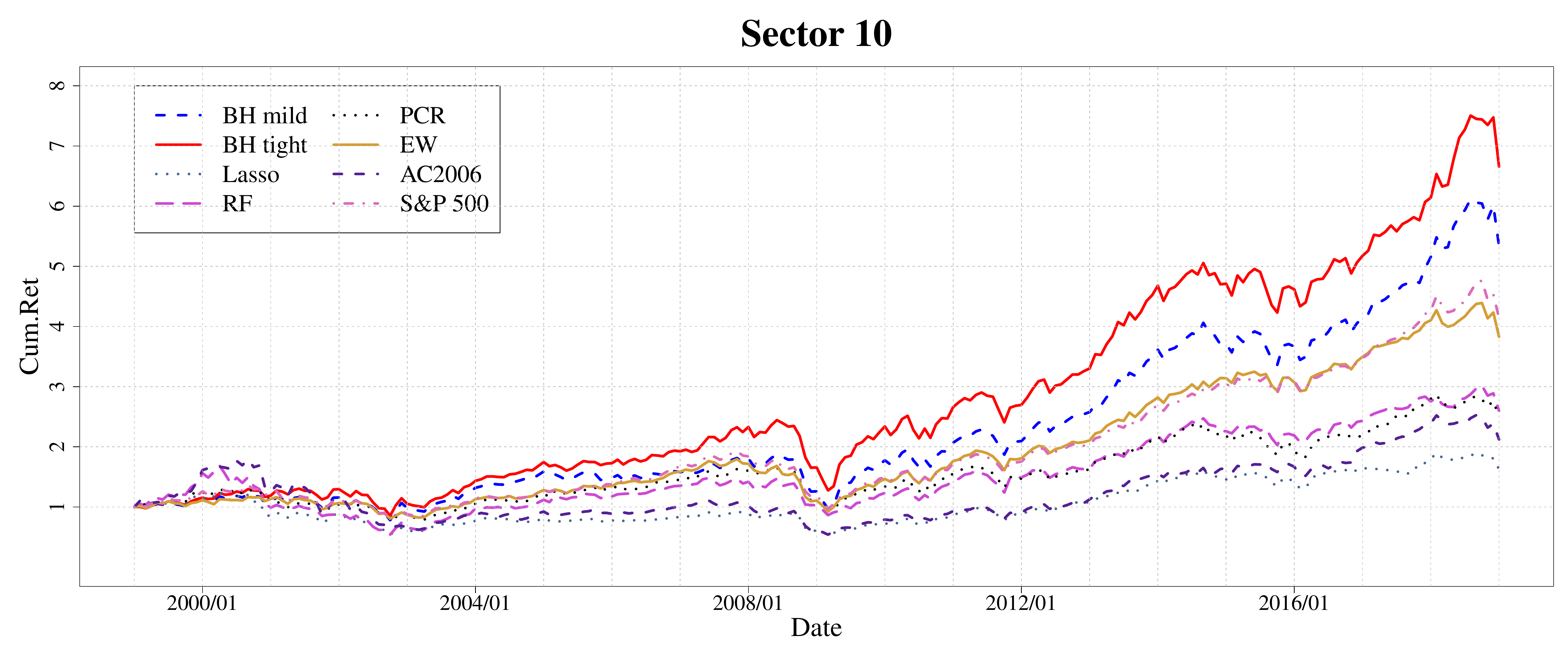}
        \vspace{0cm}
        \includegraphics[width = \textwidth]{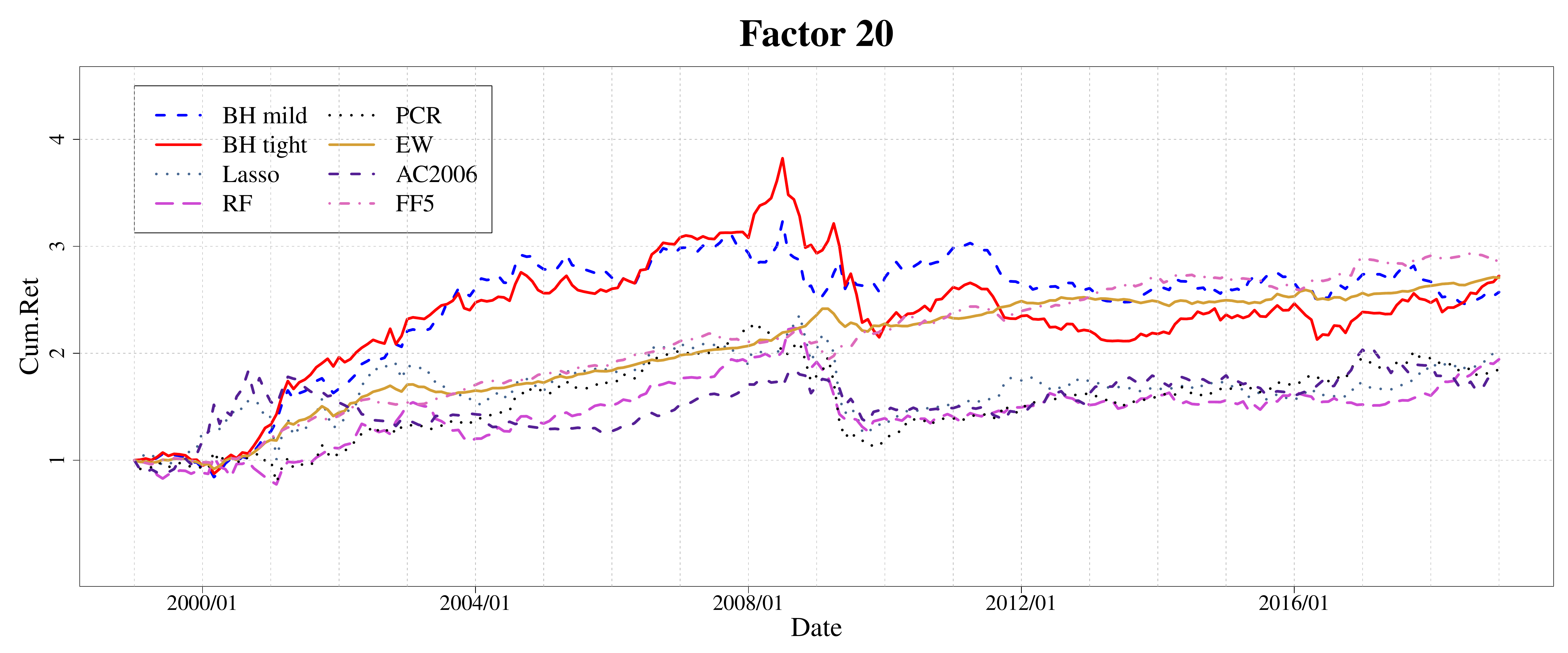}
        \vspace{-0.35cm}
        \includegraphics[width = \textwidth]{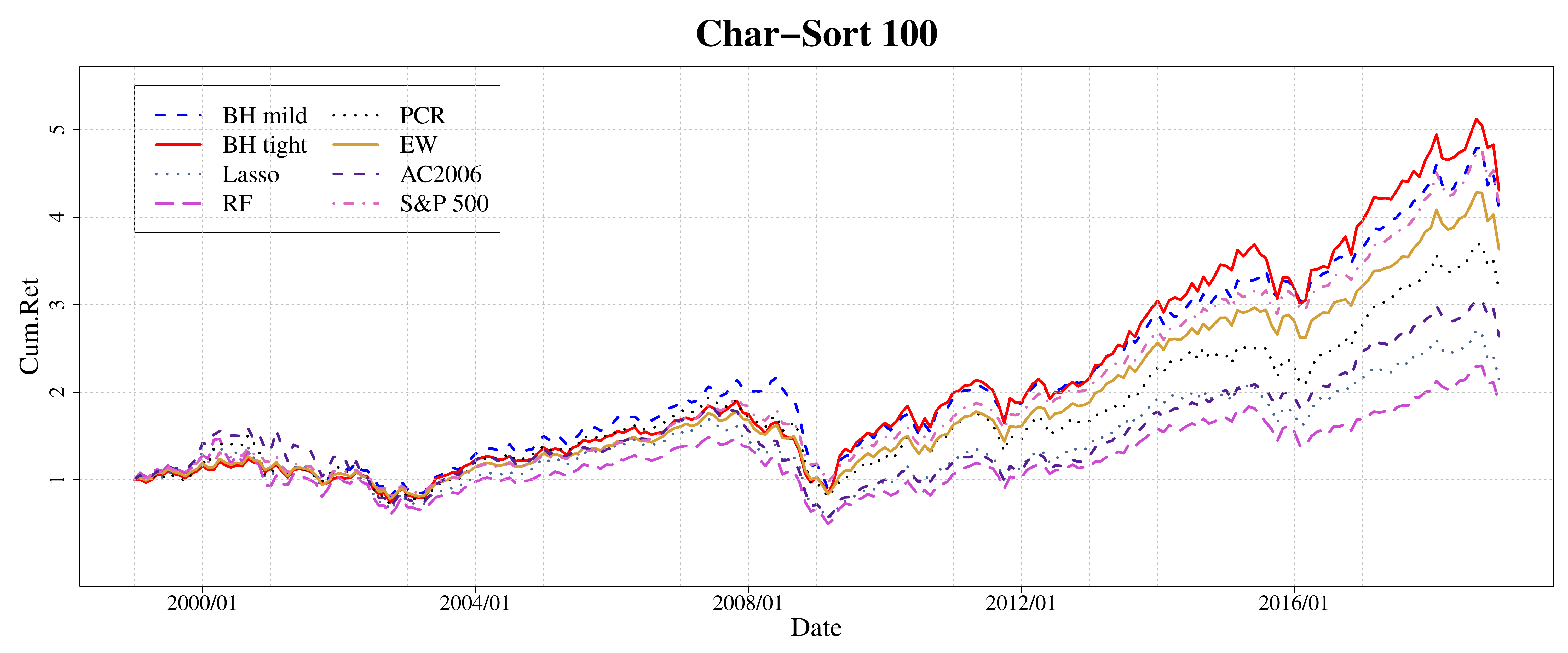}
    \end{center}
    \vspace{-0.15cm}
    \noindent \footnotesize
    This figure shows cumulative portfolio returns (compounding) for three types of portfolios and portfolio strategies in Table \ref{tab: performance}.
\end{figure}

Table \ref{tab: performance} reports portfolio performance.
All alternative methods fall behind passive investment (EW portfolio and S\&P 500) in terms of average return, annualized Sharpe ratio, and Jensen's alpha for all three groups of assets.
However, our BH approach gives the best numbers for long-only assets: 10 sector and 100 characteristics-sorted portfolios.
Figure \ref{fig: cumreturn} presents the evolution of \$1 since January 1998, where the BH approach enjoys the highest cumulative return after 20 years of investment.

\begin{figure}[ht]
    \begin{center}
        \caption{Heat Map of Weights for Sector Rotation}
        \includegraphics[width = 1\textwidth]{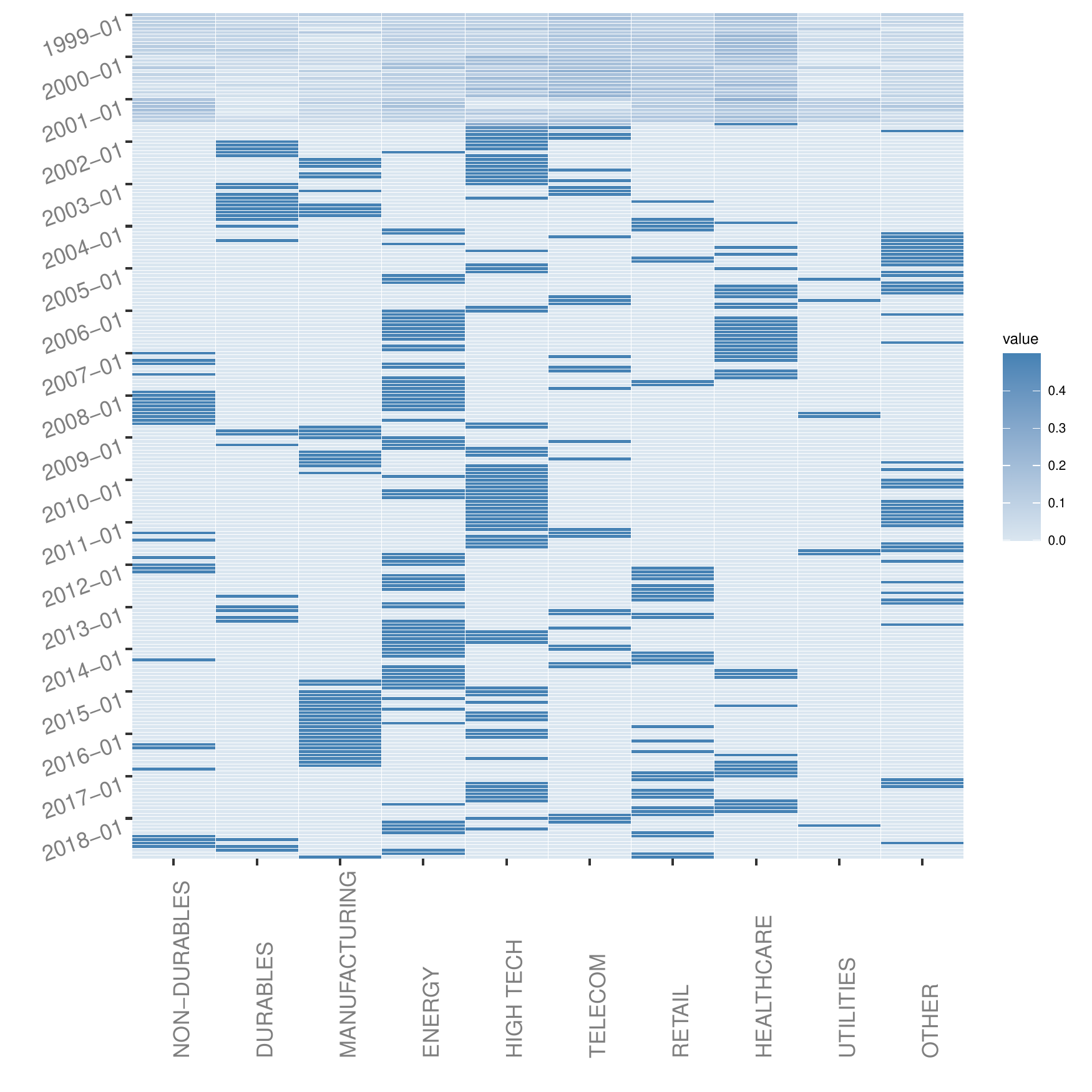}\label{fig: hm_ind}
    \end{center}
    \noindent \footnotesize
    This figure presents a heat map for dynamic portfolio weights of 10 sector portfolios in the efficient portfolio constructed by our BH (tight) approach in Table \ref{tab: performance}. We follow the Fama-French 10-industry classifications to form the 10 sector portfolios.
\end{figure}

For 10 sector portfolios, our BH approach outperforms alternatives in all categories and provides an economically and statistically significant Jensen's alpha.
The capital asset pricing model (CAPM) $R^2$ or correlation with the market factor is low, indicating the BH portfolio is an excellent diversified strategy away from the market risk.
Additionally, Figure \ref{fig: hm_ind} exhibits the changing weights over time, which tells how the 10 sectors rotate in the past 20 years.
Technology, energy, and manufacturing are the most heavily weighted sectors in the past decade.
The BH approach also obtains desirable results on 100 characteristics-sorted portfolios. It achieves the highest average returns and a Sharpe ratio similar to S\&P 500.

The 20 long-short factor investment case is interesting because investing in long-short factors is the practitioner's definition of factor investing. We also include passive investment benchmarks, such as an equally weighted portfolio using these 20 factors (EW) or Fama-French five factors. Note all approaches have flat returns after the 2008-2009 financial crisis (Figure \ref{fig: cumreturn}). Most of these 20 factors probably do not provide significant risk premia after the crisis.

\begin{figure}[ht]
    \begin{center}
        \caption{Heat Map of Weights for Factor Rotation}
        \includegraphics[width = 1\textwidth]{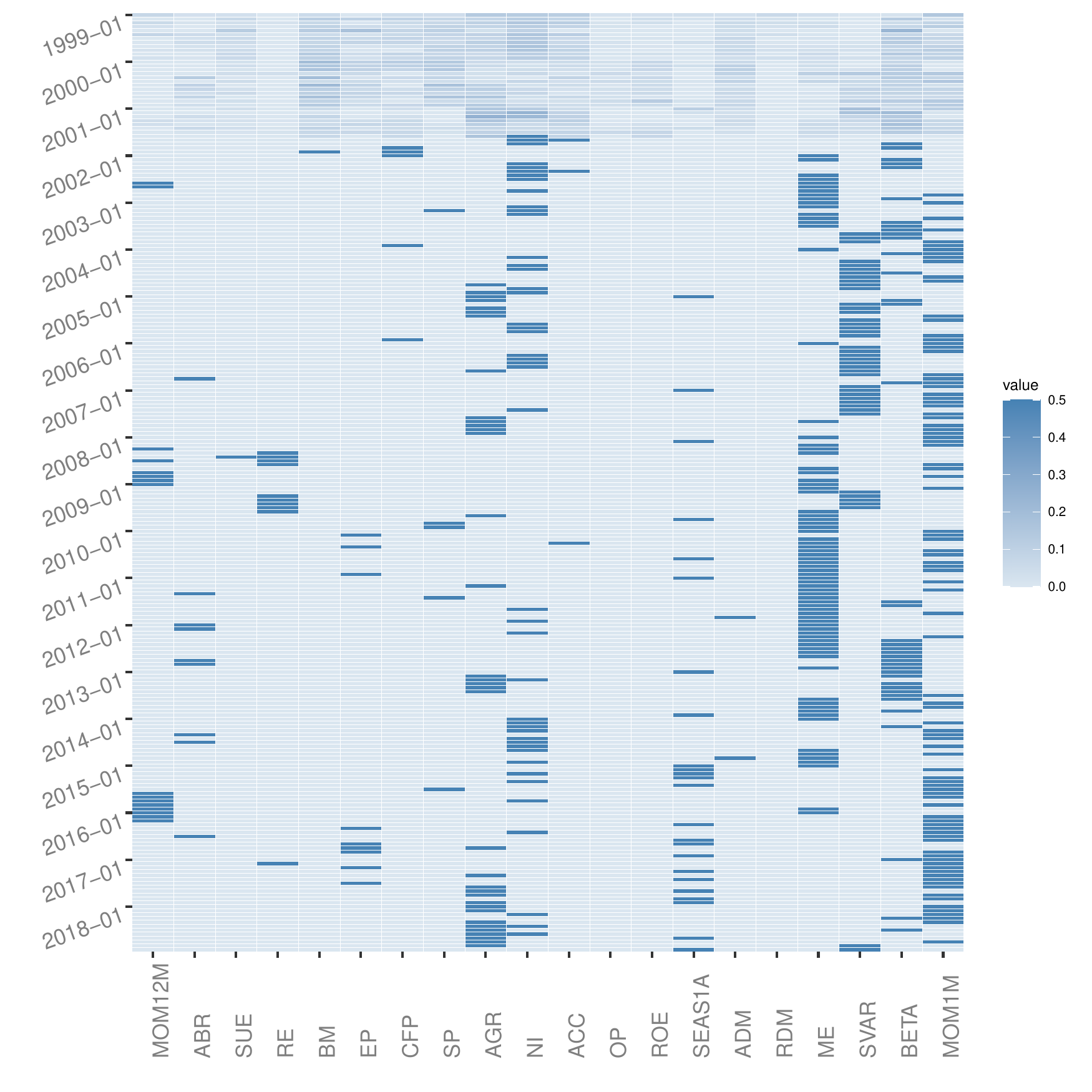}\label{fig: hm_factor}
    \end{center}
    \noindent \footnotesize This figure presents a heat map for dynamic portfolio weights of 20 long-short factors in the efficient portfolio constructed by our BH (tight) approach in Table \ref{tab: performance}. The list of risk factors are described in Appendix \ref{app: char}.
\end{figure}

An equally weighted portfolio of 20 factors has the highest Sharpe ratio, which contradicts studies that the factor zoo should be sparse \citep{feng2020taming}. Our BH approach achieves the highest monthly average return with a smaller Sharpe ratio, but the portfolio only invests a few factors in each period. Figure \ref{fig: hm_factor} shows the rotation of 20 factors over time, where market equity (size), asset growth (investment), and lag monthly return (short-term reversal) factors are heavily weighted in the past 10 years.

\subsection{Asset Pricing Implications}\label{sec: ap}
Our BH mean-variance efficient portfolio is the stochastic discount factor (SDF) for the three different cross-sections of assets. Moreover, the efficient portfolio can also be viewed as a dimension-reduced version of the factor zoo (e.g., the first principal component for the 20 factors). Therefore, this section discusses the asset pricing implications of the efficient portfolio.

First, we evaluate our BH efficient portfolio's model-adjusted performance on commonly used factor models, including CAPM, Fama-French three factors and five factors. If the efficient portfolio cannot be priced by these asset pricing models and contains additional signals, the intercept ``alpha" should be significantly positive. Table \ref{tab: ap} presents the results of the model-adjusted performance. Specifically, the BH efficient portfolio constructed by 10 sector portfolios generates significantly positive alphas over CAPM, whereas other methods fail. This monthly alphas' magnitude is economically significant, with 0.32\% and 0.29\% over CAPM and the Fama-French three-factor model. The BH efficient portfolio constructed by 20 factors has a marginally significant alpha 0.41\% over CAPM, though it shrinks to an insignificant 0.18\% over Fama-French five factors. Surprisingly, the equally weighted portfolio of 20 factors has superior performance, with a 70\% annualized Sharpe ratio and a significant alpha over Fama-French five factors.

Second, we study whether the BH efficient portfolio can explain existing anomalies. We apply our BH efficient portfolio as the SDF model to price existing anomalies: 20 published risk factors related to our characteristics. Results are summarized in Table \ref{tab: factor}.
During the past 20 years, 8 out of 20 factors have marginally significant (10\% level) alphas to CAPM and even 9 out of 20 to Fama-French three factors.
By contrast, after controlling for the BH efficient portfolio constructed by those 20 factors, only two factors still have marginally significant alphas.
We also check the first principal component of these 20 factors, which leaves six factors unexplained.
In addition, we find the maximum alpha unexplained is 0.82\% for CAPM, 0.81\% for FF3, 0.48\% for BH, and 0.63\% for PCA.
The evidence above suggests the BH SDF has the best pricing ability in this cross section of 20 long-short factors.

\section{Conclusion}  \label{sec: conclusion}

When returns are predictable, the mean-variance efficient portfolio framework's successful deployment requires estimating conditional expected returns and their covariance matrix while accounting for estimation risk.
Bayesian methods shine in these areas.
Our Bayesian hierarchical prior setting provides a way to model multiple assets other than separate time series modeling or pool modeling. It allows the sharing of information while maintaining specific heterogeneity of models for different assets. Furthermore, stock return prediction suffers from a low signal-to-noise ratio and varying predictive power over time. Our BH approach adopts heterogeneous time-varying coefficients driven by lagged fundamental characteristics to solve this problem.

Our empirical findings are also noteworthy. We study the U.S. equity market from 1978 to 2018 and find our BH approach provides better prediction performance than alternatives in terms of out-of-sample R-squared coverage of predictive intervals.
The superior coverage performance is due to our BH approach accounting for the estimation risk of regression coefficients.
Primarily for sector investing during the past 20 years, our BH approach gives average monthly returns of 0.92\% and significant Jensen`s alpha of 0.32\% .
Moreover, we study the sector and factor rotations during the past decade. We find technology, energy, and manufacturing are important sectors, and size, investment, and short-term reversal are heavily weighted factors.
In the end, the SDF constructed by our BH approach explains many risk anomalies that alternative models fail to.

\newpage
\bibliography{bayesianfinance}

\begin{thebibliography}{}

\bibitem[\protect\citeauthoryear{Amihud}{Amihud}{2002}]{amihud2002illiquidity}
Amihud, Y. (2002).
\newblock Illiquidity and stock returns: cross-section and time-series effects.
\newblock {\em Journal of financial markets\/}~{\em 5\/}(1), 31--56.

\bibitem[\protect\citeauthoryear{Avramov}{Avramov}{2004}]{avramov2004stock}
Avramov, D. (2004).
\newblock Stock return predictability and asset pricing models.
\newblock {\em The Review of Financial Studies\/}~{\em 17\/}(3), 699--738.

\bibitem[\protect\citeauthoryear{Avramov and Chordia}{Avramov and
  Chordia}{2006a}]{avramov2006asset}
Avramov, D. and T.~Chordia (2006a).
\newblock Asset pricing models and financial market anomalies.
\newblock {\em The Review of Financial Studies\/}~{\em 19\/}(3), 1001--1040.

\bibitem[\protect\citeauthoryear{Avramov and Chordia}{Avramov and
  Chordia}{2006b}]{avramov2006predicting}
Avramov, D. and T.~Chordia (2006b).
\newblock Predicting stock returns.
\newblock {\em Journal of Financial Economics\/}~{\em 82\/}(2), 387--415.

\bibitem[\protect\citeauthoryear{Avramov and Zhou}{Avramov and
  Zhou}{2010}]{avramov2010bayesian}
Avramov, D. and G.~Zhou (2010).
\newblock Bayesian portfolio analysis.
\newblock {\em Annu. Rev. Financ. Econ.\/}~{\em 2\/}(1), 25--47.

\bibitem[\protect\citeauthoryear{Barberis}{Barberis}{2000}]{barberis2000investing}
Barberis, N. (2000).
\newblock Investing for the long run when returns are predictable.
\newblock {\em The Journal of Finance\/}~{\em 55\/}(1), 225--264.

\bibitem[\protect\citeauthoryear{DeMiguel, Martín-Utrera, Nogales, and
  Uppal}{DeMiguel et~al.}{2020}]{demiguel2020portfolio}
DeMiguel, V., A.~Martín-Utrera, F.~J. Nogales, and R.~Uppal (2020).
\newblock {A transaction-cost perspective on the multitude of firm
  characteristics}.
\newblock {\em The Review of Financial Studies\/}~{\em 33\/}(5), 2180--2222.

\bibitem[\protect\citeauthoryear{Fama and French}{Fama and
  French}{1993}]{fama1993common}
Fama, E.~F. and K.~R. French (1993).
\newblock Common risk factors in the returns on stocks and bonds.
\newblock {\em Journal of Financial Economics\/}~{\em 33}, 3--56.

\bibitem[\protect\citeauthoryear{Feng, Giglio, and Xiu}{Feng
  et~al.}{2020}]{feng2020taming}
Feng, G., S.~Giglio, and D.~Xiu (2020).
\newblock Taming the factor zoo: A test of new factors.
\newblock {\em The Journal of Finance\/}~{\em 75\/}(3), 1327--1370.

\bibitem[\protect\citeauthoryear{Feng, He, He, and Polson}{Feng
  et~al.}{2020}]{feng2020deep}
Feng, G., J.~He, X.~He, and N.~Polson (2020).
\newblock Deep learning for predicting asset returns.
\newblock Technical report, City University of Hong Kong.

\bibitem[\protect\citeauthoryear{Feng, Polson, and Xu}{Feng
  et~al.}{2020}]{feng2020deepchar}
Feng, G., N.~G. Polson, and J.~Xu (2020).
\newblock Deep learning in characteristics-sorted factor models.
\newblock Technical report, City University of Hong Kong.

\bibitem[\protect\citeauthoryear{Freyberger, Neuhierl, and Weber}{Freyberger
  et~al.}{2020}]{freyberger2020dissecting}
Freyberger, J., A.~Neuhierl, and M.~Weber (2020).
\newblock Dissecting characteristics nonparametrically.
\newblock {\em The Review of Financial Studies\/}~{\em 33\/}(5), 2326--2377.

\bibitem[\protect\citeauthoryear{Fulop, Li, and Yu}{Fulop
  et~al.}{2015}]{fulop2015self}
Fulop, A., J.~Li, and J.~Yu (2015).
\newblock Self-exciting jumps, learning, and asset pricing implications.
\newblock {\em The Review of Financial Studies\/}~{\em 28\/}(3), 876--912.

\bibitem[\protect\citeauthoryear{Gu, Kelly, and Xiu}{Gu
  et~al.}{2020}]{gu2020empirical}
Gu, S., B.~Kelly, and D.~Xiu (2020).
\newblock Empirical asset pricing via machine learning.
\newblock {\em The Review of Financial Studies\/}~{\em 33\/}(5), 2223--2273.

\bibitem[\protect\citeauthoryear{Harvey and Zhou}{Harvey and
  Zhou}{1990}]{harvey1990bayesian}
Harvey, C.~R. and G.~Zhou (1990).
\newblock Bayesian inference in asset pricing tests.
\newblock {\em Journal of Financial Economics\/}~{\em 26\/}(2), 221--254.

\bibitem[\protect\citeauthoryear{Hou, Xue, and Zhang}{Hou
  et~al.}{2020}]{hou2020replicating}
Hou, K., C.~Xue, and L.~Zhang (2020).
\newblock Replicating anomalies.
\newblock {\em The Review of Financial Studies\/}~{\em 33\/}(5), 2019--2133.

\bibitem[\protect\citeauthoryear{Huang, Li, Wang, and Zhou}{Huang
  et~al.}{2020}]{huang2020time}
Huang, D., J.~Li, L.~Wang, and G.~Zhou (2020).
\newblock Time series momentum: Is it there?
\newblock {\em Journal of Financial Economics\/}~{\em 135\/}(3), 774--794.

\bibitem[\protect\citeauthoryear{Jacquier and Polson}{Jacquier and
  Polson}{2011}]{jacquier2011bayesian}
Jacquier, E. and N.~Polson (2011).
\newblock Bayesian methods in finance.
\newblock In {\em The Oxford Handbook of Bayesian Econometrics}, pp.\
  439--512.

\bibitem[\protect\citeauthoryear{Johannes, Korteweg, and Polson}{Johannes
  et~al.}{2014}]{johannes2014sequential}
Johannes, M., A.~Korteweg, and N.~Polson (2014).
\newblock Sequential learning, predictability, and optimal portfolio returns.
\newblock {\em The Journal of Finance\/}~{\em 69\/}(2), 611--644.

\bibitem[\protect\citeauthoryear{Kandel and Stambaugh}{Kandel and
  Stambaugh}{1996}]{kandel1996predictability}
Kandel, S. and R.~F. Stambaugh (1996).
\newblock On the predictability of stock returns: an asset-allocation
  perspective.
\newblock {\em The Journal of Finance\/}~{\em 51\/}(2), 385--424.

\bibitem[\protect\citeauthoryear{Kelly, Pruitt, and Su}{Kelly
  et~al.}{2019}]{kelly2019characteristics}
Kelly, B.~T., S.~Pruitt, and Y.~Su (2019).
\newblock Characteristics are covariances: A unified model of risk and return.
\newblock {\em Journal of Financial Economics\/}~{\em 134\/}(3), 501--524.

\bibitem[\protect\citeauthoryear{Kozak, Nagel, and Santosh}{Kozak
  et~al.}{2020}]{kozak2020shrinking}
Kozak, S., S.~Nagel, and S.~Santosh (2020).
\newblock Shrinking the cross-section.
\newblock {\em Journal of Financial Economics\/}~{\em 135\/}(2), 271--292.

\bibitem[\protect\citeauthoryear{Lettau and Pelger}{Lettau and
  Pelger}{2020}]{lettau2020PCA}
Lettau, M. and M.~Pelger (2020).
\newblock {Factors that fit the time series and cross-section of stock
  returns}.
\newblock {\em The Review of Financial Studies\/}~{\em 33\/}(5), 2274--2325.

\bibitem[\protect\citeauthoryear{Lewellen and Shanken}{Lewellen and
  Shanken}{2002}]{lewellen2002learning}
Lewellen, J. and J.~Shanken (2002).
\newblock Learning, asset-pricing tests, and market efficiency.
\newblock {\em The Journal of Finance\/}~{\em 57\/}(3), 1113--1145.

\bibitem[\protect\citeauthoryear{McCulloch and Rossi}{McCulloch and
  Rossi}{1991}]{mcculloch1991bayesian}
McCulloch, R. and P.~E. Rossi (1991).
\newblock A {B}ayesian approach to testing the arbitrage pricing theory.
\newblock {\em Journal of Econometrics\/}~{\em 49\/}(1-2), 141--168.

\bibitem[\protect\citeauthoryear{P{\'a}stor}{P{\'a}stor}{2000}]{pastor2000portfolio}
P{\'a}stor, L. (2000).
\newblock Portfolio selection and asset pricing models.
\newblock {\em The Journal of Finance\/}~{\em 55\/}(1), 179--223.

\bibitem[\protect\citeauthoryear{P{\'a}stor and Stambaugh}{P{\'a}stor and
  Stambaugh}{2000}]{pastor2000comparing}
P{\'a}stor, L. and R.~F. Stambaugh (2000).
\newblock Comparing asset pricing models: an investment perspective.
\newblock {\em Journal of Financial Economics\/}~{\em 56\/}(3), 335--381.

\bibitem[\protect\citeauthoryear{Plummer, Best, Cowles, and Vines}{Plummer
  et~al.}{2006}]{plummer2006coda}
Plummer, M., N.~Best, K.~Cowles, and K.~Vines (2006).
\newblock {CODA}: convergence diagnosis and output analysis for {MCMC}.
\newblock {\em R news\/}~{\em 6\/}(1), 7--11.

\bibitem[\protect\citeauthoryear{Polson and Tew}{Polson and
  Tew}{2000}]{polson2000bayesian}
Polson, N.~G. and B.~V. Tew (2000).
\newblock Bayesian portfolio selection: {A}n empirical analysis of the {S}\&{P}
  500 index 1970--1996.
\newblock {\em Journal of Business \& Economic Statistics\/}~{\em 18\/}(2),
  164--173.

\bibitem[\protect\citeauthoryear{Rossi, Allenby, and McCulloch}{Rossi
  et~al.}{2012}]{rossi2012bayesian}
Rossi, P.~E., G.~M. Allenby, and R.~McCulloch (2012).
\newblock {\em Bayesian statistics and marketing}.
\newblock John Wiley \& Sons.

\bibitem[\protect\citeauthoryear{Shanken}{Shanken}{1987}]{shanken1987bayesian}
Shanken, J. (1987).
\newblock A {B}ayesian approach to testing portfolio efficiency.
\newblock {\em Journal of Financial Economics\/}~{\em 19\/}(2), 195--215.

\bibitem[\protect\citeauthoryear{Welch and Goyal}{Welch and
  Goyal}{2008}]{welch2008comprehensive}
Welch, I. and A.~Goyal (2008).
\newblock A comprehensive look at the empirical performance of equity premium
  prediction.
\newblock {\em The Review of Financial Studies\/}~{\em 21\/}(4), 1455--1508.

\bibitem[\protect\citeauthoryear{Zellner}{Zellner}{1962}]{zellner1962efficient}
Zellner, A. (1962).
\newblock An efficient method of estimating seemingly unrelated regressions and
  tests for aggregation bias.
\newblock {\em Journal of the American statistical Association\/}~{\em
  57\/}(298), 348--368.

\end{thebibliography}

\newpage

\begin{table}[ht]
    \begin{center}
        \caption{Prediction Performance Evaluation} \label{tab: prediction}
        \begin{tabular}{c|c|rrrrrr}
            \toprule
            Assets                         & Metrics     & BH tight & BH mild & Lasso  & RF     & PCR    & AC2006 \\
            \midrule
            \multicolumn{8}{c}{Overall, 1999 - 2018} \\
            \midrule
            \multirow{3}{*}{Sector 10}     & $R_{OOS}^2$ & -0.007   & -0.010  & -0.018 & -0.040 & -0.053 & -0.139 \\
                                           & $IL_{OOS}$  & 0.277    & 0.290   & 0.206  & 0.173  & 0.200  & 0.285  \\
                                           & $C_{OOS}$   & 94.5\%   & 93.6\%  & 93.5\% & 89.1\% & 92.2\% & 92.3\% \\
            \midrule
            \multirow{3}{*}{Factor 20}     & $R_{OOS}^2$ & -0.026   & -0.028  & -0.013 & -0.048 & -0.075 & -0.165 \\
                                           & $IL_{OOS}$  & 0.236    & 0.244   & 0.157  & 0.127  & 0.152  & 0.255  \\
                                           & $C_{OOS}$   & 95.4\%   & 95.6\%  & 91.5\% & 86.8\% & 90.6\% & 92.6\% \\
            \midrule
            \multirow{3}{*}{Char-Sort 100} & $R_{OOS}^2$ & -0.022   & -0.002  & -0.014 & -0.041 & -0.060 & -0.119 \\
                                           & $IL_{OOS}$  & 0.267    & 0.294   & 0.189  & 0.158  & 0.184  & 0.285  \\
                                           & $C_{OOS}$   & 95.1\%   & 94.3\%  & 93.4\% & 88.9\% & 92.4\% & 92.4\% \\
            \midrule
            \multicolumn{8}{c}{1999 - 2008} \\
            \midrule
            \multirow{3}{*}{Sector 10}     & $R_{OOS}^2$ & -0.027   & -0.010  & -0.038 & -0.041 & -0.055 & -0.155 \\
                                           & $IL_{OOS}$  & 0.275    & 0.283   & 0.205  & 0.172  & 0.199  & 0.284  \\
                                           & $C_{OOS}$   & 93.3\%   & 93.6\%  & 91.2\% & 87.0\% & 89.9\% & 91.6\% \\
            \midrule
            \multirow{3}{*}{Factor 20}     & $R_{OOS}^2$ & -0.025   & -0.027  & -0.015 & -0.053 & -0.074 & -0.171 \\
                                           & $IL_{OOS}$  & 0.225    & 0.233   & 0.147  & 0.119  & 0.143  & 0.245  \\
                                           & $C_{OOS}$   & 93.8\%   & 94.0\%  & 85.7\% & 79.0\% & 84.3\% & 93.4\% \\
            \midrule
            \multirow{3}{*}{Char-Sort 100} & $R_{OOS}^2$ & -0.036   & -0.015  & -0.034 & -0.054 & -0.056 & -0.127 \\
                                           & $IL_{OOS}$  & 0.263    & 0.292   & 0.187  & 0.156  & 0.182  & 0.279  \\
                                           & $C_{OOS}$   & 95.0\%   & 94.0\%  & 92.3\% & 87.2\% & 91.2\% & 91.0\% \\
            \midrule
            \multicolumn{8}{c}{2009 - 2018} \\
            \midrule
            \multirow{3}{*}{Sector 10}     & $R_{OOS}^2$ & 0.006    & -0.010  & 0.007  & -0.031 & -0.053 & -0.138 \\
                                           & $IL_{OOS}$  & 0.279    & 0.296   & 0.207  & 0.173  & 0.200  & 0.285  \\
                                           & $C_{OOS}$   & 95.7\%   & 93.6\%  & 95.8\% & 91.2\% & 94.5\% & 93.1\% \\
            \midrule
            \multirow{3}{*}{Factor 20}     & $R_{OOS}^2$ & -0.026   & -0.029  & -0.008 & -0.032 & -0.074 & -0.131 \\
                                           & $IL_{OOS}$  & 0.247    & 0.255   & 0.166  & 0.134  & 0.161  & 0.264  \\
                                           & $C_{OOS}$   & 97.0\%   & 97.2\%  & 97.3\% & 94.7\% & 96.8\% & 91.8\% \\
            \midrule
            \multirow{3}{*}{Char-Sort 100} & $R_{OOS}^2$ & 0.001    & -0.001  & 0.001  & -0.026 & -0.066 & -0.119 \\
                                           & $IL_{OOS}$  & 0.271    & 0.296   & 0.192  & 0.160  & 0.186  & 0.291  \\
                                           & $C_{OOS}$   & 95.2\%   & 94.6\%  & 94.5\% & 90.5\% & 93.6\% & 93.9\% \\
            \bottomrule
        \end{tabular}
    \end{center}
    \noindent \footnotesize
    This table presents the prediction performance of different portfolio strategies from 1999 to 2018. Three panels represent the overall, first, and second of sample periods. In each panel, the out-of-sample $R^2$ (with respect to the moving average), 95\% prediction interval length, and coverage are reported for different portfolio strategies (BH, Lasso, random forest, principal component regression, and the Bayesian method of \cite{avramov2006predicting}). We show two types of hyperparameters for the BH method for robustness.
\end{table}

\newpage
\begin{table}[ht]
    \begin{center}
        \caption{Risk-Adjusted Portfolio Performance} \label{tab: performance}
        \begin{tabular}{cccccccccc}
            \toprule
            \multicolumn{9}{c}{Sector 10}\\
            \midrule
                     & BH tight & BH mild & Lasso   & RF      & PCR     & EW     & AC2006  & S\&P 500 \\
            \midrule
            Avg.     & 0.92\%   & 0.82\%  & 0.33\%  & 0.54\%  & 0.48\%  & 0.64\% & 0.61\%  & 0.68\%   \\

            SR       & 0.53     & 0.47    & 0.13    & 0.26    & 0.26    & 0.43   & 0.30    & 0.45     \\
            $\alpha$ & 0.32\%   & 0.23\%  & -0.26\% & -0.08\% & -0.05\% & 0.10\% & -0.02\% & -0.03\%  \\
            tstat    & 2.01     & 1.43    & -1.67   & -0.48   & -0.36   & 1.40   & -0.09   & -0.67    \\
            $R^2$    & 0.76     & 0.75    & 0.76    & 0.76    & 0.73    & 0.93   & 0.74    & 0.97     \\
            \toprule
            \multicolumn{9}{c}{Factor 20}\\
            \midrule
                     & BH tight & BH mild & Lasso   & RF      & PCR     & EW     & AC2006  & FF5      \\
            \midrule
            Avg.     & 0.47\%   & 0.44\%  & 0.40\%  & 0.34\%  & 0.34\%  & 0.43\% & 0.31\%  & 0.44\%   \\
            SR       & 0.35     & 0.35    & 0.20    & 0.19    & 0.17    & 0.70   & 0.15    & 0.74     \\
            $\alpha$ & 0.41\%   & 0.35\%  & 0.36\%  & 0.21\%  & 0.31\%  & 0.35\% & 0.15\%  & 0.10\%   \\
            tstat    & 2.03     & 1.91    & 1.29    & 0.88    & 1.18    & 4.60   & 0.62    & 1.16     \\
            $R^2$    & 0.06     & 0.04    & 0.05    & 0.00    & 0.07    & 0.27   & 0.00    & 0.16     \\

            \toprule
            \multicolumn{9}{c}{Char-Sort 100}\\
            \midrule
                     & BH tight & BH mild & Lasso   & RF      & PCR     & EW     & AC2006  & S\&P 500 \\

            \midrule
            Avg.     & 0.75\%   & 0.71\%  & 0.49\%  & 0.47\%  & 0.64\%  & 0.64\% & 0.59\%  & 0.68\%   \\
            SR       & 0.39     & 0.40    & 0.21    & 0.18    & 0.31    & 0.38   & 0.25    & 0.45     \\
            $\alpha$ & 0.10\%   & 0.09\%  & -0.21\% & -0.28\% & -0.03\% & 0.04\% & -0.12\% & -0.03\%  \\
            tstat    & 0.71     & 0.82    & -1.56   & -1.56   & -0.19   & 0.95   & -0.76   & -0.67    \\
            $R^2$    & 0.85     & 0.89    & 0.87    & 0.86    & 0.84    & 0.98   & 0.83    & 0.97
            \\\bottomrule
        \end{tabular}
    \end{center}
    \noindent \footnotesize
    This table presents the risk-adjusted performance of different portfolio strategies from 1999 to 2018. Three panels represent three types of portfolios. In each panel, the average monthly return, annualized Sharpe ratio, Jensen's alpha, t-statistics, and CAPM $R^2$ are reported for different portfolio strategies (BH, Lasso, random forest, principal component regression, and the Bayesian method of \cite{avramov2006predicting}). We show two types of hyperparameters for the BH method for robustness.
\end{table}

\begin{table}[htbp]
    \begin{center}
        \caption{Model-Adjusted Portfolio Performance}\label{tab: ap}
        \begin{tabular}{cccccccc}
            \toprule
                                  &          & BH tight & Lasso   & RF      & PCR     & EW      & AC2006  \\
            \midrule
            \multicolumn{8}{c}{Sector 10} \\
            \midrule
            \multirow{2}{*}{CAPM} & $\alpha$ & 0.32\%   & -0.26\% & -0.08\% & -0.05\% & 0.10\%  & -0.02\% \\
                                  & tstat    & 2.01     & -1.67   & -0.48   & -0.36   & 1.40    & -0.09   \\
            \midrule
            \multirow{2}{*}{FF3}  & $\alpha$ & 0.29\%   & -0.24\% & -0.04\% & -0.09\% & 0.08\%  & -0.03\% \\
                                  & tstat    & 1.88     & -1.53   & -0.24   & -0.57   & 1.47    & -0.15   \\
            \midrule
            \multirow{2}{*}{FF5}  & $\alpha$ & 0.23\%   & -0.15\% & -0.01\% & -0.12\% & -0.05\% & 0.20\%  \\
                                  & tstat    & 1.42     & -0.95   & -0.05   & -0.82   & -0.99   & 1.14    \\

            \midrule
            \multicolumn{8}{c}{Factor 20} \\
            \midrule
            \multirow{2}{*}{CAPM} & $\alpha$ & 0.41\%   & 0.36\%  & 0.21\%  & 0.31\%  & 0.35\%  & 0.15\%  \\
                                  & tstat    & 2.03     & 1.29    & 0.88    & 1.18    & 4.60    & 0.62    \\
            \midrule
            \multirow{2}{*}{FF3}  & $\alpha$ & 0.36\%   & 0.34\%  & 0.17\%  & 0.21\%  & 0.34\%  & 0.07\%  \\
                                  & tstat    & 1.86     & 1.21    & 0.74    & 0.84    & 4.92    & 0.30    \\

            \midrule
            \multirow{2}{*}{FF5}  & $\alpha$ & 0.18\%   & 0.18\%  & 0.14\%  & 0.16\%  & 0.14\%  & 0.23\%  \\
                                  & tstat    & 0.88     & 0.63    & 0.60    & 0.64    & 2.58    & 0.98    \\

            \midrule
            \multicolumn{8}{c}{Char-Sort 100} \\
            \midrule
            \multirow{2}{*}{CAPM} & $\alpha$ & 0.10\%   & -0.21\% & -0.28\% & -0.03\% & 0.04\%  & -0.12\% \\
                                  & tstat    & 0.71     & -1.56   & -1.78   & -0.19   & 0.95    & -0.76   \\
            \midrule
            \multirow{2}{*}{FF3}  & $\alpha$ & 0.05\%   & -0.27\% & -0.27\% & -0.08\% & 0.01\%  & -0.17\% \\
                                  & tstat    & 0.35     & -2.08   & -2.08   & -0.57   & 0.45    & -1.04   \\

            \midrule
            \multirow{2}{*}{FF5}  & $\alpha$ & 0.05\%   & -0.14\% & -0.16\% & 0.13\%  & -0.01\% & 0.08\%  \\
                                  & tstat    & 0.42     & -1.03   & -1.14   & 0.99    & -0.45   & 0.50    \\

            \bottomrule
        \end{tabular}
    \end{center}
    \noindent \footnotesize
    This table presents the model-adjusted performance ``alpha" with respect to different asset pricing factor models from 1999 to 2018. Three panels represent three types of portfolios. In each panel, the intercept ``alpha" and t-statistics are reported for different portfolio strategies (BH, Lasso, random forest, principal component regression, equally weighted portfolio, and the Bayesian method of \cite{avramov2006predicting}) and different benchmark asset pricing factor models (CAPM, Fama-French 3 and 5 factors).
\end{table}

\begin{table}[ht]
    \begin{center}
        \caption{BH Efficient Portfolio: Dissecting Anomalies}\label{tab: factor}
        \begin{tabular}{r|rr|rr|rr|rr}
            \toprule
            & \multicolumn{2}{c|}{CAPM} & \multicolumn{2}{c|}{FF3} & \multicolumn{2}{c|}{BH (Factor 20)} & \multicolumn{2}{c}{PCA (Factor 20)}  \\
            \midrule
                   & $\alpha$ & tstat & $\alpha$ & tstat & $\alpha$ & tstat & $\alpha$ & tstat \\
            \hline
            MOM12M & 0.53\%   & 1.28  & 0.56\%   & 1.42  & 0.30\%   & 0.70  & 0.18\%   & 0.54  \\
            ABR    & 0.14\%   & 0.83  & 0.16\%   & 1.00  & 0.11\%   & 0.66  & 0.05\%   & 0.32  \\
            SUE    & 0.40\%   & 2.18  & 0.42\%   & 2.47  & 0.29\%   & 1.58  & 0.28\%   & 1.70  \\
            RE     & 0.41\%   & 1.45  & 0.55\%   & 2.06  & 0.15\%   & 0.50  & 0.11\%   & 0.57  \\
            BM     & -0.00\%  & -0.02 & -0.21\%  & -0.89 & 0.04\%   & 0.12  & 0.23\%   & 0.84  \\
            EP     & 0.82\%   & 2.70  & 0.81\%   & 3.50  & 0.32\%   & 1.09  & 0.49\%   & 1.72  \\
            CFP    & 0.44\%   & 1.75  & 0.35\%   & 1.67  & 0.21\%   & 0.89  & 0.37\%   & 1.47  \\
            SP     & 0.48\%   & 1.69  & 0.27\%   & 1.22  & 0.48\%   & 1.70  & 0.63\%   & 2.33  \\
            AGR    & 0.18\%   & 0.79  & 0.02\%   & 0.12  & 0.06\%   & 0.30  & 0.16\%   & 0.72  \\
            NI     & 0.55\%   & 2.79  & 0.54\%   & 3.30  & 0.22\%   & 1.02  & 0.29\%   & 1.59  \\
            ACC    & -0.01\%  & -0.08 & 0.04\%   & 0.36  & 0.14\%   & 0.79  & 0.06\%   & 0.31  \\
            OP     & 0.56\%   & 2.79  & 0.68\%   & 4.06  & 0.22\%   & 0.92  & 0.23\%   & 1.79  \\
            ROE    & 0.62\%   & 2.80  & 0.73\%   & 3.82  & 0.24\%   & 0.90  & 0.23\%   & 1.88  \\
            SEAS1A & -0.05\%  & -0.17 & 0.05\%   & 0.21  & 0.06\%   & 0.24  & -0.00\%  & -0.01 \\
            ADM    & 0.37\%   & 1.95  & 0.28\%   & 1.70  & 0.23\%   & 1.21  & 0.30\%   & 1.55  \\
            RDM    & 0.17\%   & 0.75  & 0.24\%   & 1.27  & 0.36\%   & 1.60  & 0.39\%   & 1.36  \\
            ME     & 0.32\%   & 1.25  & 0.03\%   & 0.25  & 0.47\%   & 1.71  & 0.56\%   & 2.86  \\
            SVAR   & 0.56\%   & 1.57  & 0.68\%   & 2.34  & -0.29\%  & -0.63 & -0.21\%  & -0.99 \\
            BETA   & 0.45\%   & 1.64  & 0.41\%   & 1.53  & 0.21\%   & 0.85  & 0.34\%   & 1.34  \\
            MOM1M  & 0.16\%   & 0.50  & 0.11\%   & 0.34  & 0.20\%   & 0.62  & 0.36\%   & 1.19  \\
            \bottomrule
        \end{tabular}
    \end{center}
    \noindent \footnotesize
    This table presents the asset pricing implication of our BH efficient portfolios. The first two columns show the CAPM and Fama-French 3-factor alphas and their t-statistics with respect to 20 long-short factors from 1999 to 2018. The factor description is listed in Appendix \ref{app: char}. The third column reports the fitted alphas and t-statistics for our BH efficient portfolio, and the last one for the first principal component.
\end{table}
\newpage

\newpage

\newpage

\clearpage
\begin{appendix}
    \section{Equity Characteristics}\label{app: char}

    Following \cite{hou2020replicating}, we categorize the equity characteristics into six groups. The data are available on their website \href{http://global-q.org/testingportfolios.html}{http://global-q.org/testingportfolios.htm}.
    We replicate their downloadable factors successfully.
    Then, we follow their equity characteristics formulas and create 20 representative characteristics for our data library.
    All these characteristics data are available since January 1978.

    For our research, we need data that can be updated every month. However, many of these characteristics calculations involve quarterly or even annually updated financial variables. Therefore, we need some adjustments to make them update monthly, and our data updating principle is straightforward. We update those formula variables as late as possible. For example, corporate earnings are updated quarterly from the corporate earnings report, and the share price can be updated monthly (or at a higher frequency). We calculate the earnings-to-price ratio by updating the nominator quarterly and the denominator monthly.

    Finally, we take a six-month lag for using the data for accounting information from the annual report. We use their annually updated accounting variables at the end of June in the second year for most companies. And for quarterly accounting information, we take a three-month lag. Though conservative, these procedures are standard for using data in Compustat.

    \subsection{Momentum}

    Cumulative abnormal returns around earnings announcement dates (ABR)

    Standard Unexpected Earnings (SUE).

    Revisions in analyst earnings forecasts (RE).

    Cumulative Returns on prior 2-12 month (MOM12M).

    \subsection{Value versus growth}

    Book-to-Market (BM)

    Earnings-to-Price (EP)

    Cashflow-to-Price (CFP)

    Sales-to-Price (SP)

    \subsection{Investment}

    Asset Growth Rate (AGR)

    Net Equity Issuance (NI, or share repurchase)

    Accruals (ACC)

    \subsection{Profitability}

    Operating Profitability (OP)

    Return on Equity (ROE)

    \subsection{Intangibles}

    Seasonality (SEAS1A)

    Advertisement-to-Market (ADM)

    Research and Design Expense to market (RDM)

    \subsection{Frictions (or Size)}

    Market Equity (ME)

    Stock Variance (SVAR)

    CAPM Beta (BETA)

    Short-term Reversal (MOM1M)

    \section{Market-Timing Macro Predictors} \label{app: macro}
    The market timing macro predictors used in this paper is as follows.
    \subsection{Equity Market Aggregate Variables}

    We calculate the following equity characteristics for individual stocks in S\&P 500 and aggregate them as the value-weighted average to proxy for macro predictors. The equity characteristics include Dividend Yield (MKTDY), Earnings-to-Price (MKTEP), Book-to-Market (MKTBM), Net Equity Issuance (MKTNI), and Stock Variance (MKTSVAR).

    \subsection{Amihud Illiquidity (ILL)}

    Our AMILL is the amihud illiquidity of the S\&P 500 index, a proxy for the market illiquidity. Following \cite{amihud2002illiquidity}, we calculate the monthly illiquidity measure of individual equities using the daily returns and trading volumes over one month. We cover the constituents of the S\&P 500 index, and calculate the value-weighted average of the individual equities as the illiquidity of S\&P 500 index.

    \subsection{3-Month Treasury Bills (TBL)}

    Following \cite{welch2008comprehensive}, the TBL is the 3-Month Treasury Bill : Secondary Market Rate from the economic research database at \href{https://fred.stlouisfed.org/series/DTB3}{the Federal Reserve Bank at St. Louis}.

    \subsection{Inflation (INFL)}

    Following \cite{welch2008comprehensive}, the INFL is the \textit{Consumer Price Index (All Urban Consumers)} from the Bureau of Labor Statistics.

    \subsection{Default Yield Spread (DFY)}

    Following \cite{welch2008comprehensive}, the DFY is the difference between BAA- and AAA-rated corporate bond yields.

    \subsection{Term Spread (TMS)}

    Following \cite{welch2008comprehensive}, the TMS is the difference between the long-term yield on government bonds (by Ibbotson Associates) and the Treasury bill.

\clearpage
    \section{MCMC Convergence}\label{app: mcmc}
    In this section, we demonstrate the MCMC results and convergence performance. The quality of posterior distribution draws is one primary concern for all MCMC simulation studies. In particular, we check the convergence performance of the MCMC for four parameters $\{b, \Sigma, \bar{b}, \delta_b\}$ because they are the key parameters for the Bayesian hierarchical structure. We show their trace plots and effective sample size (ESS).

\begin{figure}[ht]
        \begin{center}
            \caption{Trace Plot of Posterior Draws for One Randomly Selected Element of $\{b, \Sigma, \bar{b}, \delta_b\}$.}
            \includegraphics[width=0.85\textwidth]{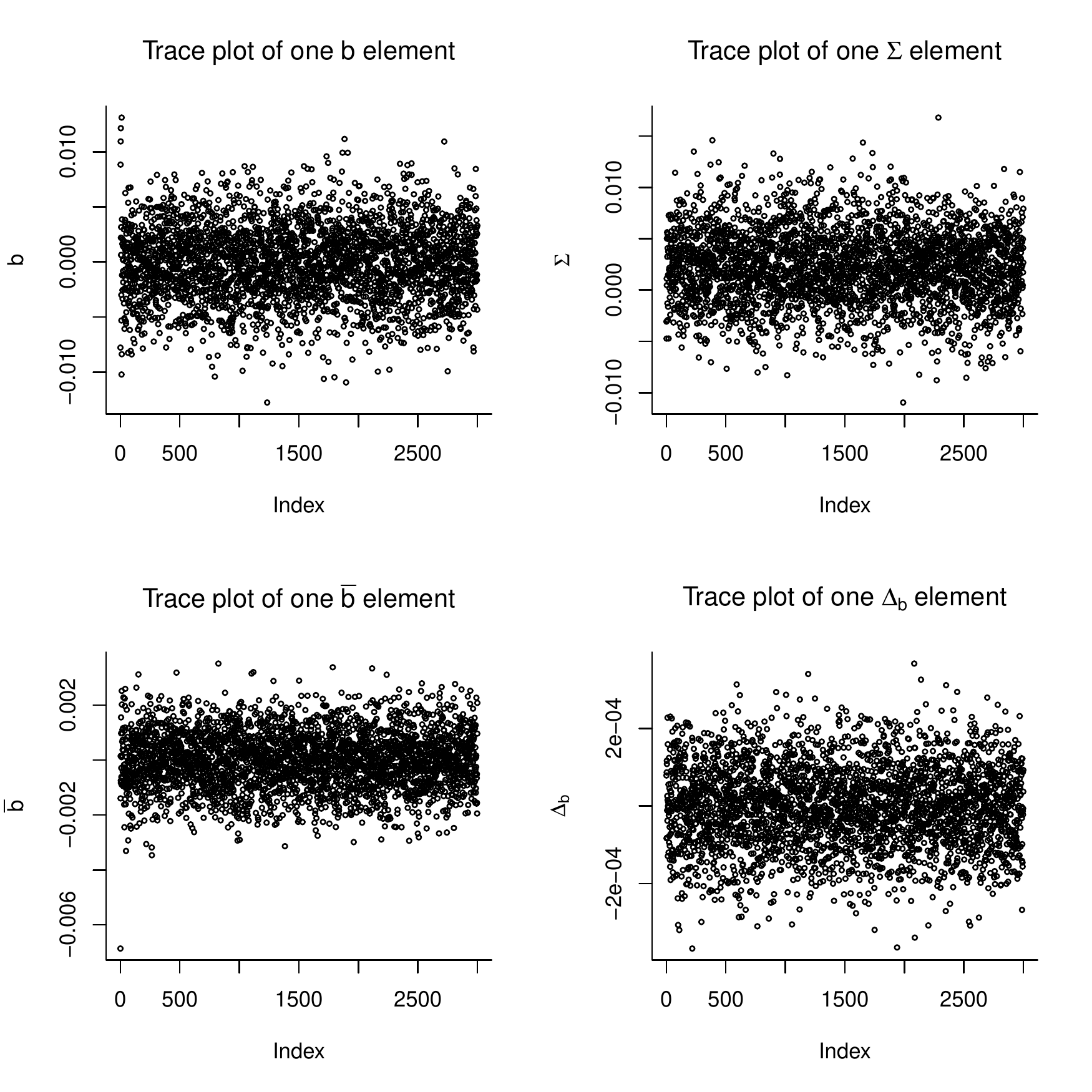} \label{plot:trace}
        \end{center}
        \noindent \footnotesize
This figure shows that the posterior draws stabilize quickly. Since $\{b, \Sigma, \bar{b}, \delta_b\}$ are multi-dimensional, we only show trace plot of one randomly chosen elements of those parameters. All other parameters shows similar pattern.
    \end{figure}

\begin{figure}[ht]
        \begin{center}
            \caption{Histogram Plots for Ratios of ESS/Sample Size for All Elements of $\{b, \Sigma, \bar{b}, \delta_b\}$.}
            \includegraphics[width=0.85\textwidth]{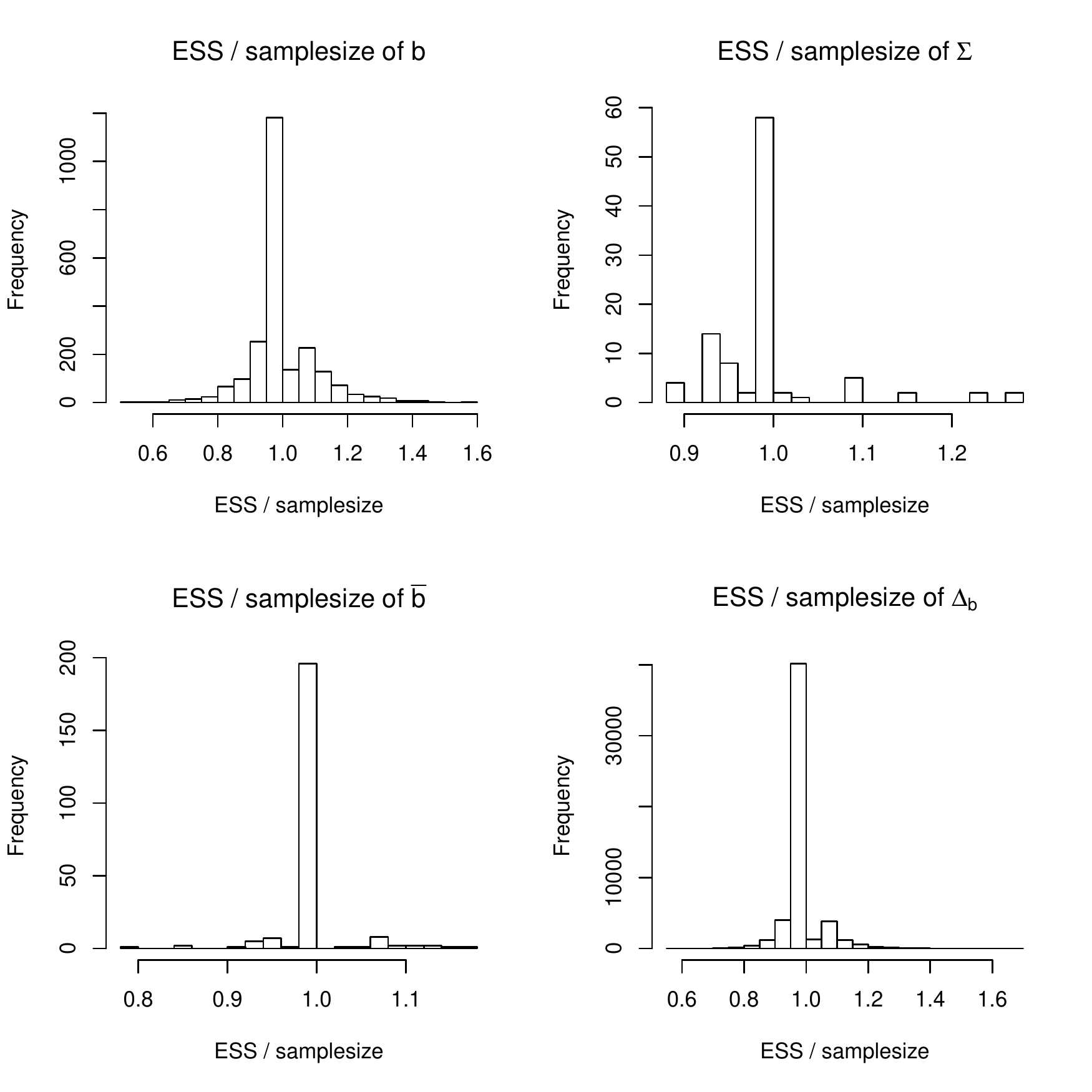} \label{plot:ESS}
        \end{center}
        \noindent \footnotesize
For each element of $\{b, \Sigma, \bar{b}, \delta_b\}$, we calculate the ratio of effective sample size to the total number of posterior draws, then draw a histogram of those ratios. From the plots we can see that the minimal ratio is greater than 0.6, whereas most of them are around 1, which indicates that the effective sample size is very high, and the sampler explores the posterior distribution efficiently.
    \end{figure}

    Due to the high-dimensional nature of the parameters, showing the trace plots of all elements is impossible. Therefore, we present the trace plot of one randomly selected element of each parameter in Figure \ref{plot:trace}. Additionally, we calculate the effective sample size of all elements using the R package \texttt{coda} from \cite{plummer2006coda} and plot the histogram for the ratio of the effective sample size to the total number of posterior draws on all elements in Figure \ref{plot:ESS}. In our study, the algorithm draws 3,000 posterior samples.

    Figure \ref{plot:trace} shows that the Markov chains for $\{b, \Sigma, \bar{b}, \delta_b\}$ stabilize quickly following inception. Figure \ref{plot:ESS} confirms this viewpoint because all ESS/Sample Size ratios are greater than 0.6, whereas most of them are around 1. Note that when the auto-correlation of a Markov chain is low, \texttt{coda} package may calculate the effective sample size as higher than the actual number of posterior draws.

\end{appendix}

\end{document}